\documentclass[%
aip,
amsmath,amssymb,
superscriptaddress,
 preprint,%
]{revtex4-2}

\usepackage[utf8]{inputenc}
\usepackage{amsmath,amssymb,amsfonts,bm}
\usepackage{graphicx,epstopdf}  
\usepackage{color}
\usepackage{graphicx}
\usepackage{float}
\usepackage{physics}
\usepackage{braket,   cleveref}

\newcommand{\B}{\mbox{\tiny B}}

\newcommand{\M}{\mbox{\tiny M}}

\newcommand{\tS}{\mbox{\tiny S}}
\newcommand{\I}{\mbox{\tiny I}}

\newcommand{\la}{\langle}
\newcommand{\ra}{\rangle}

\newcommand{\nl}{\nonumber \\}
\newcommand{\be}{\begin{equation}}
\newcommand{\ee}{ \end{equation}}
\newcommand{\bsube}{\begin{subequations}}
\newcommand{\esube}{\end{subequations}}
\newcommand{\Eq}[1]{Eq.\,(\ref{#1})}

\newcommand{\Fig}[1]{Fig.\,\ref{#1}}

\newcommand{\RN}[1]{%
  \textup{\uppercase\expandafter{\romannumeral#1}}%
}

\newcommand{\bea}{\begin{eqnarray}}
	\newcommand{\eea}{\end{eqnarray}}
\newcommand{\ba}{\begin{array}}
	\newcommand{\ea}{\end{array}}

\newcommand{\bl}{\begin{flalign}}
	\newcommand{\enl}{\end{flalign}}

\newcommand{\mc}[1]{\mathcal{#1}}

\newcommand{\tdse}{time-dependent Schr\"{o}dinger equation}

\newcommand{\eq}[1]{Eq. \eqref{#1}}

\newcommand{\half}{\frac{1}{2}}

\renewcommand{\bf}[1]{\mathbf{#1}}

\begin{document}

\title{Geometric phase-induced nuclear quantum interference is robust against quantum dissipation}


\author{Xiang Li}

\author{Bing Gu}
\email{gubing@weatlake.edu.cn}

\affiliation{Institute of Natural Sciences,  
Westlake Institute for Advanced Study,   Hangzhou,   Zhejiang 310024,   China}

\affiliation{Department of Chemistry and Department of Physics,   Westlake University,   Hangzhou,   Zhejiang 310030,   China}

\date{\today}
\begin{abstract}
One of the intriguing effects due to conical intersections is the geometric phase,   manifested as destructive quantum interference  in the nuclear probability distribution.  However,   whether such geometric phase-induced interference can survive in dissipative environments remains an open question. We demonstrate by  numerically exact dissipative conical intersection dynamics simulations that the destructive interference is highly robust against non-Markovian quantum dissipation. To do so,   we integrate the recently proposed local diabatic representation to describe vibronic couplings  and the hierarchical equations of motion for system-bath interactions.
Both vibrational and electronic environments are considered. 
An intuitive path integral-like picture is provided to explain the robustness of geometric phase-induced interference.
\end{abstract}

\maketitle

\section{Introduction}
Conical intersections associated with electronic degeneracy play critical roles in photochemistry and photophysics  \cite{domcke2012, koppel2004, larson2020,  levine2007, mead1979, mead1992, joubert-doriol2013, bersuker2006, gu2020b, gu2020c, larson2020, domcke2011,prigogine2002,mead1982, 
		casida2012, polli2010, mandal2023, gu2020a, rafiq2023, quenneville2003} . 
 One of the intriguing effects due to conical intersections (CIs) is the geometric phase. 
For time-reversal-invariant electronic Hamiltonian, the adiabatic electronic wavefunction will acquire a phase of $\pi$  upon traversing a closed loop around a CI  \citep{berry1979, berry1984}. As the total molecular wavefunction is single-valued, the multivaluedness of the adiabatic wavefunction is compensated by the same phase in the nuclear wavefunction.    
The geometric phase  is originated from the nontrivial geometry of the adiabatic electronic state space \cite{mead1992, xie2025.11124} and often manifests as a destructive interference pattern as e.g. a nodal line in the nuclear wave packet dynamics \citep{MeadRevModPhys.64.51}.  
The geometric phase can have a significant influence on reaction pathways,   product distributions,   and vibrational spectrum \cite{wittig2012, 
yuan2018a,  ryabinkin2017}.
It should be emphasized that the geometric phase is only meaningful using the adiabatic electronic states (or more precisely, a geometrically nontrival molecular fiber bundle). If a crude adiabatic representation is employed, there will be no geometric phase. 

  Chemical reactions rarely occur in isolation but are subject to dissipative interactions with environments such as solvents or interfaces.  Environmental dissipation can strongly alter the 
 reaction dynamics by  quantum dissipation and decoherence \cite{Yan885160,nitzan_2024}. 
 Exploring how the  environment influce nonadiabatic dynamics near CIs is important for understanding reactions in complex envrionments.  
 Previous studies have focused on the reaction rates and  population dynamics   \cite{KUHL2000227,   C6FD00088F,Tan1222A550,Duan16382,IKEDA2018203}, whether the geometric phase-induced interference in the wavepacket dynamics can survive in dissipative environments remains an open question. 

In this article,   we investigate nuclear wave packet dynamics near CIs under  non-Markovian quantum dissipation by an numerically exact modeling of dissipative conical intersection dynamics method. We focus on how quantum decoherence and dissipation influence  geometric phase, or more precisely, the geometric phase-induced quantum interference. Theoretically,   we employ Feynman path integrals and influence functional theory  \citep{Fey63118} to dissect the impact of dissipation on geometric-phase-induced wavepacket interference. 
We combine the recently proposed local diabatic representation (LDR) to describe vibronic couplings  \cite{gu2023,   gu2024a} and the hierarchical equations of motion (HEOM) method to describe the system-bath coupling  \cite{tanimura1989, tanimura1990, Xu05041103,  Xu07031107, jin2008,yan2014,Yan04216}  
for the numerically exact modeling of dissipative conical intersection dynamics. 

With a two-dimensional vibronic coupling model   \cite{xie2017b} coupled to either an electronic or a vibrational bath,   we demonstrate that despite dissipation alters the nonadiabatic transitions of the electrons and the density distribution of the nuclears,    the destructive interference pattern induced by geometric phase remains robust against strong dissipation.

This paper is organized as follows: In Sec.II, we present the theories of LDR,  Feynman-Vernon path integral and HEOM. In Sec. III, we investigate the dynamics of nuclear wavepacket and electron population with the vibrational bath and the electronic bath via the LDR--HEOM method.  The following discussion about the effect of quantum dissipation is presented based on the Feynman--Vernon path integral theory.  We summarize the paper in Sec. IV. 

In this article,   we use atomic units with $k_\text{B} =\hbar=1$.

\section{Formalism} \

\subsection{Local diabatic representation of conical intersection dynamics} \label{sec:theory}

The most widely used framework for nonadiabatic molecular dynamics is based on the Born-Huang expansion,    \be
	\Psi({\bf r},  {\bf R},  t)=\sum_\alpha \phi_\alpha({\bf r};{\bf R})\chi_\alpha({\bf R},  t).
\ee  
where  $\phi_\alpha({\bf r}; {\bf R})$ is the adiabatic electronic states which depends parametrically
on the nuclear configuration $\bf R$,   i.e.
\begin{align}
	H_{\rm BO}({\bf R})|\phi_\alpha ({\bf R}) \ra= V_{\alpha} ({\bf R})| \phi_\alpha({\bf R})\ra
\end{align}
where $H_{\rm BO}({\bf R}) = H_\text{M} - \hat{T}_\text{N}$
is the electronic Hamiltonian,   the full
molecular Hamiltonian subtracting the nuclear kinetic energy
operator,    $V_{\alpha}({\bf R})$ is the $\alpha$th adiabatic PES,   and $\chi_\alpha (R,   t)$ is the nuclear wave packet evolving on the $\alpha$th adiabatic PES. 
However,   in this Born-Huang representation,   the non-Born-Oppenheimer effects including geometric phase and nonadiabatic transitions,   critical for conical intersection dynamics,   are accounted for by singular terms such as the Berry connection and derivative couplings. 
To remove the divergences,    we recently proposed a local diabatic representation (LDR)  \cite{gu2023,   gu2024a}.  
In it,    the ansatz for the full molecular wavefunction is  given by
\begin{align}\label{LDR}
\Psi(\bf r,   \bf R,   t) &= \sum_{\bf n} {\sum_\alpha C_{\bf n\alpha}(t) \phi_\alpha(\bf r; \bf R_{\bf n})} \chi_{\bf n}(\bf R) 
\nl
&\equiv \sum_{\bf n \alpha} C_{{\bf n} \alpha}(t) | {{\bf n} \alpha}\ra
\end{align}
where $\phi_\alpha(\bf R_{\bf n})$ is  the $\alpha$th adiabatic electronic eigenstate of the electronic Born-Oppenheimer Hamiltonian at the nuclear geometry $\bf R_{\bf n} $ with energy $V_\alpha(\bf R_{\bf n})$,   
$\chi_{\bf n}({\bf R})$ is the orthonormal discrete variable representation nuclear basis for the nuclear wavefunction,   localized at ${\bf R}_{\bf n}$.

Inserting \Eq{LDR} into the molecular \tdse\ $i \partial {\Psi(\bf r,   \bf R,   t)} /\partial t=H\Psi(\bf r,   \bf R,   t)$ with 
the molecular  Hamiltonian 
$
H = \hat{T}_\text{N} + H_\text{BO}(\bf r; \bf R),  
$
 and left multiply $\bra{\bf m \beta}$ yields the equation of motion for the expansion coefficients

\be
i \dot{C}_{\bf m \beta}(t)
= V_{\bf m \beta}  C_{\bf m \beta}(t)+  \sum_{\bf n,   \alpha} T_{\bf m \bf n}A_{\bf m \beta,   \bf n \alpha} C_{\bf n\alpha}(t)
\label{eq:main}.
\ee
Here
$ T_{\bf m \bf n} = \la{\chi_{\bf m}| \hat{T}_\text{N} | \chi_{\bf n}}\ra_{\bf R} 
$ is the kinetic energy operator matrix elements and the electronic overlap matrix 
\be
A_{\bf m\beta,   \bf n \alpha} = \la{\phi_{\beta}(\bf R_{\bf m}) | \phi_{\alpha}(\bf R_{\bf n}) }\ra_{\bf r},  
\ee
where $\la{\cdots}\ra_{\bf r}$ ($\la{\cdots}\ra_{\bf R}$) denotes the integration over electronic (nuclear) degrees of freedom.
In deriving \eq{eq:main},   we have made use of 
$ H_\text{BO}(\bf R) |{{\bf n} \alpha} \ra \approx V_\alpha(\bf R_{\bf n}) |{{\bf n} \alpha}\ra  
$
as the nuclear state $|{\bf n}\ra$ is an eigenstate of all position operators. 
Although we have employed the adiabatic electronic states,   \eq{eq:main} does not contain any singularities because the nuclear kinetic energy operator does not operate on the electronic states. Therefore,   in contrast to wavepacket dynamics in the Born-Huang representation,   the singularity of the derivative couplings at the CI  will not affect the simulations based on the LDR. 

The geometrical information of the vibronic Hilbert space is encoded in the global overlap matrix $A_{{\bf m} \beta,   {\bf n}\alpha}$. 
Specifically,   the geometric phase around the CI can be obtained from the product of overlap matrices along a closed loop  
\begin{align}
e^{i\theta_\alpha}= &\lim_{N \rightarrow \infty} 	\la{\phi_{\alpha}(\bf R_{ 1}) | \phi_{\alpha}(\bf R_{\bf 2}) }\ra_{\bf r} \la{\phi_{\alpha}(\bf R_{ 2}) | \phi_{\alpha}(\bf R_{3}) }\ra_{\bf r}\la \phi_{\alpha}({\bf R}_3)|
\nl
& \cdots | \phi_{\alpha}(\bf R_{ N}) \ra_{\bf r} \la{\phi_{\alpha}(\bf R_{ N}) | \phi_{\alpha}(\bf R_{ 1}) }\ra_{\bf r}
\nl
=& \lim_{N \rightarrow \infty}  A_{1\alpha,   2\alpha}A_{2\alpha,   3\alpha} \cdots A_{{N-1}\alpha,   {N} \alpha} A_{{N}\alpha,   1\alpha}
\end{align}
If the loop encircles a CI,   $\theta_\alpha = \pi$. 
For time-reversal symmetric Hamiltonian,   the geometric phase $\theta_\alpha \in \set{0,   \pi}$ is topological as it is invariant under any local changes of the loop.


In the LDR,   the time-evolution operator is given by 
\begin{align}
&{G}({\bf R}_f,   \alpha_f,   t;  {\bf R}_0,   \alpha_0,  t_0)\equiv \la {\bf R}_f \alpha_f | e^{-iHt}| {\bf R}_0 \alpha_0 \ra
\nl
&=\sum_{\{\alpha_j,   {\bf R}_j \}} \prod_{j=0}^{L-1} \la {\bf R}_{j+1} \alpha_{j+1}| e^{-i(\hat{T}_{\rm N}+\hat{H}_{\rm BO}) \Delta t} |{\bf R}_{j} \alpha_j\ra
\nl
&=\sum_{\{\xi(t) \}} W[\xi(t)] e^{iS[\xi(t)]}
\end{align}
where $\xi(t) = \set{\xi_0,   \xi_1,   \cdots,   \xi_L}$ represents all discrete paths in the joint electronic-nuclear space satisfying the boundary conditions $\xi_0 = (\bf R_0,   \alpha_0),   \xi_L = (\bf R_f,   \alpha_f)$. Each $\xi_j = (\bf R_j,   \alpha_j)$ labels both the nuclear configuration and electronic state at time $t_j = j \Delta t$.
Here the action
\begin{align}
S[\xi(t)]=-\sum_{j=0}^{N_t } \Big(K_{{\bf R}_{j+1},  {\bf R}_{j}} +
  V_{\alpha_j}({\bf R}_j)\Big) \Delta t,  
\end{align}
with $K_{{\bf R}_{j+1},  {\bf R}_{j}}=\ln(\la {\bf R}_{j+1} | e^{-i\hat{T}_{\rm N} \Delta t}| {\bf R}_{j}\ra)/\Delta t$   and the geometric factor 
\begin{align}
W[\xi(t)]=\prod_{j } A_{\xi_{j+1},  \xi_{j}}.
\end{align}
accounts for the geometric phase effects and nonadiabatic transitions.

The path integral-like picture is particularly convenient to understand how geometric phase  can induce destructive nuclear quantum interference around a conical intersection.
 Consider two adiabatic paths $\Gamma_-$ and $\Gamma_+$  surrounding a CI.  When the dynamical actions of two paths are equal ${S}[\Gamma_-]={S}[\Gamma_+] = S $,   the sum of the transition amplitudes is 
 \be 
\mc{A}(\bf R_f,   \bf R_0) = e^{iS} ( W[\Gamma_-]+W[\Gamma_+] ) = e^{iS} W[\Gamma_-](1+e^{i \theta_{\Gamma}})=0.
 \ee 
 where $\Gamma$ refers to the closed loop traversing along $\Gamma_+$ first and then along $\Gamma_-$ backwards. 
 
 
 \begin{figure}[htbp]
 	\centering
 	\includegraphics[width=0.4\textwidth]{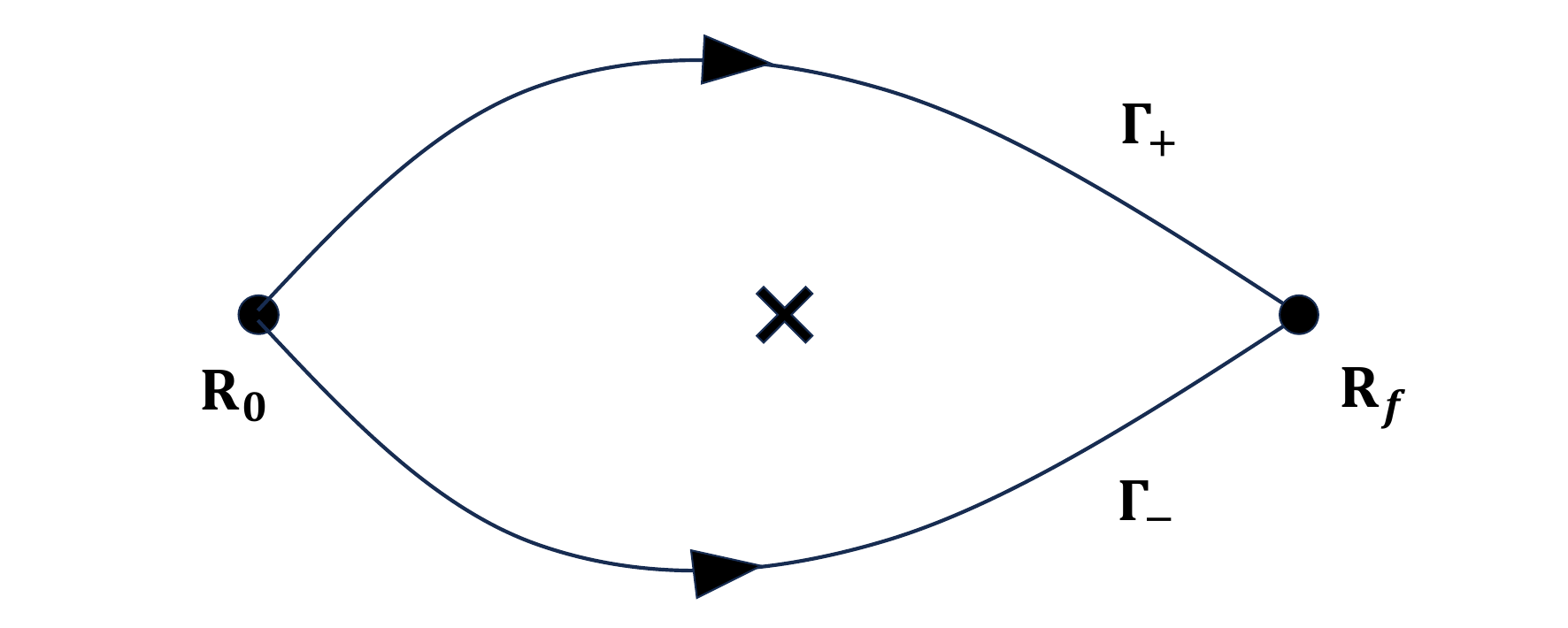}
 	\caption{Schematic illustration of two paths $ {\Gamma_+,   \Gamma_-} $ connecting  ${\bf R}_0$ and ${\bf R}_f$ and  surrounding the CI (marked by $\times$) }
 	\label{fig_path}
 \end{figure}
 
\subsection{Influence functional for dissipative conical intersection dynamics}
We employ the influence functional to develop an intuitive picture of how a dissipative environment can influence the conical intersection dynamics.  
We consider  a bosonic Gaussian bath,   which is widely used for modeling the solvent environments.
It contains noninteracting bosons and have linear coupling with the system. The total Hamiltonian is
$
  H= H_{\text{M}}+H_{\text{I}} +H_{\text{B}},
$
where  $H_\text{B}$ is the bath Hamiltonian. The system-bath coupling is assumed to be 
\begin{align}
  H_{\rm I}=Q \otimes X.
\end{align}
where $Q$ is an arbitrary Hermitian operator acting on either nuclear or electronic space,   $X$ is a collective bath coordinate,   which is a summation of microscopic degrees of freedom.
 The  bath influence on the system dynamics is completely characterized by the bath correlation function
\begin{align}
  C(t)=\la X(t) X(0) \ra_\text{B},  
\end{align}
with $X(t)=e^{-iH_{\B}t}Xe^{iH_{\B}t}$ and
$\la \cdots \ra_{\B}={\rm tr}_{\B}\{\cdots \rho^{\rm eq}_{\B}\}$ for the canonical state $\rho^{\rm eq}_\text{B}=e^{-\beta H_\text{B}}/Z$.
The bath correlation function can be obtained from the spectral density $J(\omega)$ via the fluctuation-dissipation theorem,  
\begin{align}\label{FDT}
C(t)
  =\frac{1}{\pi}
      \!\int^{\infty}_{-\infty}\!\!{\rm d}\omega\,  
        \frac{e^{-i\omega t}J(\omega)}{1-e^{-\beta\omega}}.
\end{align}
with inverse temperature $\beta = 1/T$. Here the spectral density,  reads
\begin{align}\label{JB}
  J(\omega)\equiv\frac{1}{2}\!\int^{\infty}_{-\infty}\!\!{\rm d}t\,  
    e^{i\omega t} \la[ X(t),   X(0)]\ra_{\B}.
\end{align}
For Gaussian bath,   $J(\omega)$ is  a coupling strength-weighted density of states,    independent with the state of bath.

In the LDR the vibronic density operator is represented by
\begin{align} 
	\hat{\rho}(t) =\sum_{{\bf n}{\bf n'},  {\alpha \alpha'}}\rho_{{\bf n}\alpha,  {{\bf n'} \alpha'}}(t) |{\bf n}\alpha\ra \la {\bf n'}{\alpha'}|
	= \sum_{\xi,  \xi'}\rho_{\M} (\xi,  \xi',  t) |\xi\ra \la\xi'|
\end{align}
To lighten the notation,   we introduce the composite index $\xi = \set{\bf n,   \alpha}$. 
The evolution of reduced density operator can be represented as
\begin{align}
 {\rho}(\xi,  \xi',   t)=\sum_{\xi_0,  \xi'_0}  \mathcal{G}(\xi,  \xi',  t;\xi_0,  \xi_0',  t_0)
  {\rho}_{\M}(\xi_0,  \xi'_0,   t_0)
\end{align}
Here the propagator $\mathcal{G}(\xi,  \xi',  t;\xi_0,  \xi_0',  t_0)$ can be obtained from the path integral.
\begin{align}
&\mathcal{G}(\xi,  \xi',  t;\xi_0,  \xi_0',  t_0)=
\nl
&\sum_{\{\xi(t),  \xi'(t)\}} W[\xi(t) ] e^{iS[\xi(t)]} \mathcal{F}[Q(\xi(t)),  Q(\xi'(t))] \overline{W}[\xi'(t)] e^{-iS[\xi'(t)]}
\end{align}
$Q[\xi_j]$ is the trajectory of  $Q$,   depending on the composite path $\xi_j$. 
$\mathcal{F}[Q(\xi(t)),  Q(\xi'(t))]$ is the  influence functional which represents the bath's influence on the evolution of the molecular system \citep{Fey63118}.  For Gaussian bath,   the influence functional is given by 
\begin{align}
  \mathcal{F}[Q(\xi(t)),  Q(\xi'(t))] =\exp\Big\{&-\sum_j^{N_t} (Q[\xi_j] - Q[\xi'_j] ) \Delta t
\nl
&\sum^j_l  \big[ C(t_j-t_l)Q[\xi_l]-C^*(t_j-t_l)Q[\xi'_l] \big]\Delta t \Big\}
\end{align}

Although Markovian quantum master equations such as Lindblad and Redfield equations are widely used for open quantum dynamics  \cite{Wei21015008},   the exact reduced system dynamics  is non-Markovian,   meaning that the time-evolution is not only decided by the current state $\rho(t)$,   but also by its history $\rho(t'),   t' < t$. The memory length is characterized by the bath correlation function $C(t)$.

 \begin{figure*}[htbp]
  \centering
  \includegraphics[width=0.75\textwidth]{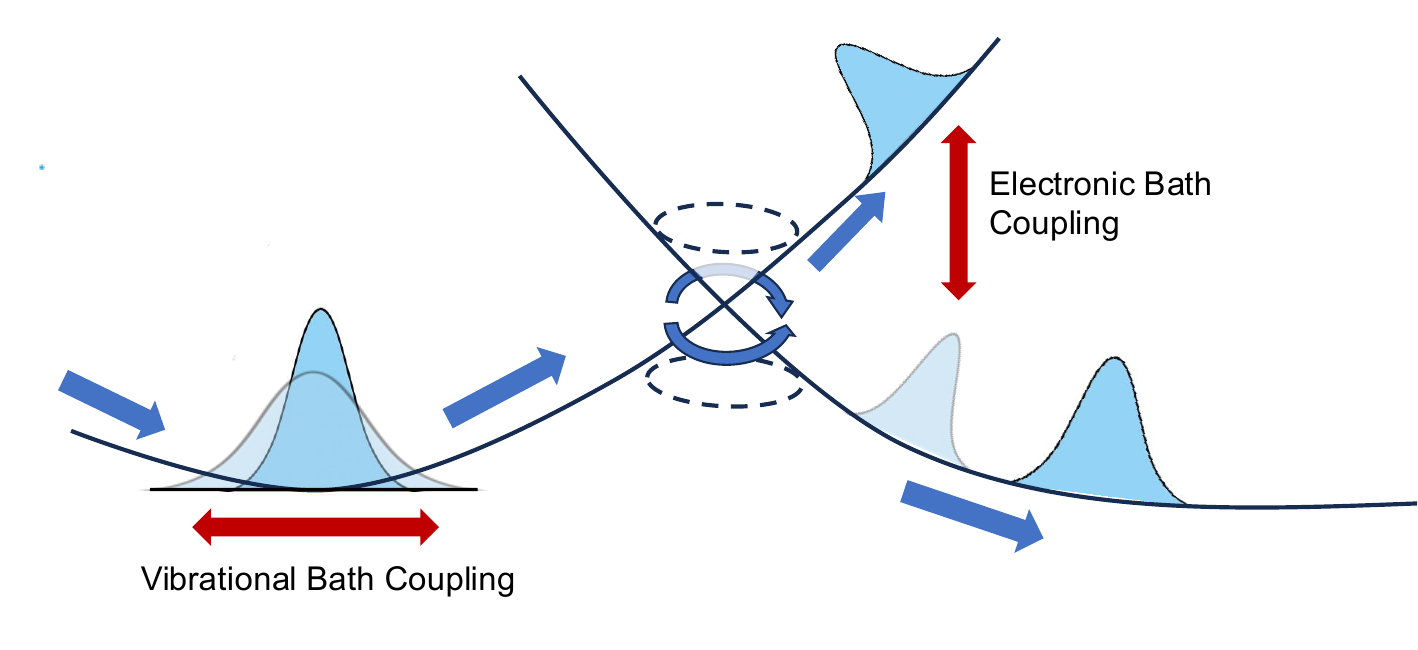}
  \caption{Schematic illustration of nonadiabatic molecular dynamics near CI with vibrational bath coupling and electronic bath coupling}
  \label{demo}
\end{figure*}

\subsection{Hierarchical equations of motion} 
We employ the hierarchical equations of motion (HEOM),   an exact and non-perturbative method for the simulation of non-Markovian open quantum dynamics  \cite{tanimura1989, tanimura1990, Xu05041103,  Xu07031107, jin2008,yan2014,Yan04216, tanimura2020,wang2022e}.  In HEOM,   the bath correlation function is decomposed by a sum of exponentials
\be \label{FBt_corr}
\la X^{\B}(t)X^{\B}(0)\ra_{\B}
=\sum^K_{k=1}\eta_k e^{-\gamma_k t}.
\ee
with the time reversal symmetry of the correlation function $C(-t)= C^* (t)$,  
\be \label{FBt_corr_rev}
\la X^{\B}(0)X^{\B}(t)\ra_{\B}
=\sum^{K}_{k=1}\eta_{k}^{\ast} e^{-\gamma_k^{\ast} t}\equiv\sum^{K}_{k=1} \eta_{\bar k}^{\ast} e^{-\gamma_k t}.
\ee
Such expansion
can be achieved by  a  sum--over--poles expression
for the Fourier integrand on the right--hand--side of \Eq{FDT},  
followed by the Cauchy's contour integration \cite{Hu10101106,  Hu11244106,  Din11164107,  Din12224103,  Zhe121129}
,   or  using the time--domain fitting decomposition
scheme  \cite{Che22221102,  Tak24204105}.
The second equality of \Eq{FBt_corr_rev} is due to the fact that
the exponents in \Eq{FBt_corr} 
must be either real or complex conjugate paired  \cite{Che22221102},   and thus
we may determine $\bar k$ in the index set $ \{k=1,  2,  ...,  K\}$
by the pairwise equality $\gamma_{\bar k}=\gamma_{k}^{\ast}$ \cite{Yan16110306}.
Via above exponential decomposition we can obtain the HEOM from the Feynman-Vernon influence functional:
\begin{align}\label{DEOM}
  \dot\rho^{(L)}_{\bf L}=&
  -i[{H}_\text{M},  \rho^{(L)}_{\bf L}]-\sum_k L_k \gamma_{k}\rho^{(L)}_{\bf L}
   -i\sum_{k}[{Q},  \rho^{(L+1)}_{{\bf L}_{k}^+}]
   \nl
 &-i\sum_{k}L_{k}\big(\eta_{k}{Q}\rho^{(L-1)}_{{\bf L}_{k}^-}
    -\eta_{\bar k}^{\ast}\rho^{(L-1)}_{{\bf L}_{k}^-}{Q}\big).
 \end{align}
Here we denote ${\bm \ell}=[\ell_1,  \cdots,  \ell_K]$ as the levels of the hierarchy and $L=\sum_k \ell_k$ with $\ell_k=0,  1,  2,  \cdots$. ${\bm \ell}^{\pm}_k$ means the $[L_1,  \cdots,  L_k \pm 1,  \cdots,  L_K]$. The time-dependence of the density operators is suppressed in \cref{DEOM}. 
The zeroth-tier $\rho^{(L)}_{\bf L}$ is  the molecular reduced density matrix $\rho^{(0)}_{\bf 0}={\rho}$. The higher-order  $\{\rho^{(L)}_{\bf L}\}$
are auxiliary density operators (ADOs),    which contain the information of the history of evolution and the system-bath correlation \cite{tanimura2006,  Zhu12194106, wang2022e, Li_2023}.  
Each $\rho^{(L)}_{\bf L}$ connects with higher-tier ADOs $\rho^{(L+1)}_{{\bf L}_{k}^+}$ and lower-tier ADOs $\rho^{(L-1)}_{{\bf L}_{k}^-}$,   forming the hierarchical structure.
 As $\{\ell_k\}$ increase,    the damping terms  $\{-\sum_k \ell_k \gamma_{k} \rho^{(L)}_{\bm \ell}\}$ will suppress the high-order ADOs,    which allows us to truncate the HEOM at an enough high tier of the hierarchy \cite{Zha15214112,  Zha23014106}.

\section{Results and Discussion}

We  apply the LDR-HEOM method to simulate the dissipative conical intersection dynamics of 
\begin{figure}[t]
  \centering
  \includegraphics[width=0.5\textwidth]{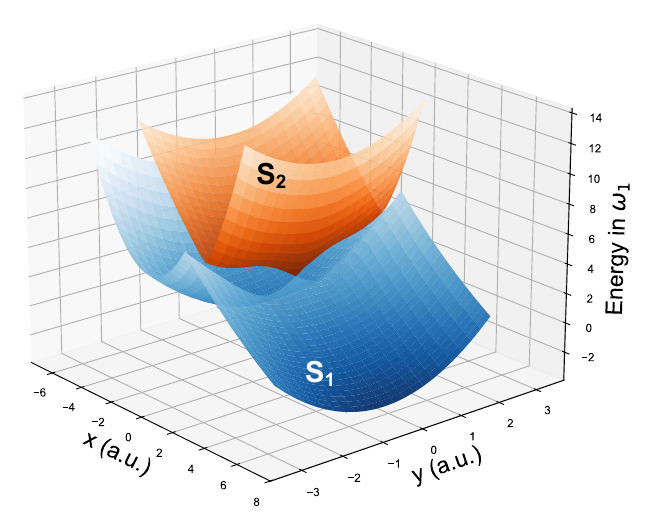}
  \caption{Adiabatic potential energy surfaces of the vibronic coupling model.  
  The CI is located at $(0.275,   0)$. }
  \label{fig_PES}
\end{figure}
a two-state,   two-dimensional vibronic coupling model  \cite{xie2017b},    resembling the photodissociation of phenol,   with the Hamiltonian given by
\be
H_{\M} = \hat{T}_\text{N} \bf I + \bf V
\ee
where $\bf I$ is the identity matrix in the electronic space,   and the nuclear kinetic energy operator  $\hat{T}_\text{N} =- \half ({{\partial}^2/{\partial x^2}+{\partial}^2/{\partial y^2}})$. 
The diabatic potential energy matrix $\bf V$ consists of  diabatic potential energy surfaces \cite{xie2017b}
\begin{align}
\begin{split}
V_{11} &= \frac{\omega^2_1}{2} \qty(x+\frac{a}{2})^2 +  \frac{\omega^2_2}{2} y^2 
\\
V_{22} &= Ae^{-(x+b)/D} + \frac{\omega^2_2}{2} y^2 - \Delta
\end{split}
\end{align}
and the  diabatic coupling  
$
V_{12}=V_{21}=cy e^{-(x-x_\text{CI})^2/{2\sigma^2_x} - y^2/2\sigma^2_y }
$. 
Here,   $x$ resembles the stretching of the O-H bond and  $y$ represents the coupling mode.  
The diabatic coupling is linear around the CI and damped by a Gaussian function away from it. 
The model parameters are (in a.u.)  $\omega_1 = \omega_2 = 1$,   $a=4$,   $b=-11$,   $c=2$,   $A=5$,   $\Delta=12$,   $x_\text{CI}=0$,   $D=10$,   $\sigma_x=1.699$,   and $\sigma_y=0.849$.

The adiabatic potential energy surfaces (\Fig{fig_PES}) ${\rm S}_1$ and ${\rm S}_2$,   obtained by diagonalizing the diabatic potential energy matrix,   show an energetically inaccessible conical intersection flanked by two energetically lower saddle points. The CI is located at $(x,   y) = (0.275,   0)$ with energy $E_\text{CI} = 2.586$. The energy of the two equivalent saddle points is 1.854,   forming a potential barrier along the tuning mode.


We consider  two types of dissipation,   as illustrated in \Fig{demo}: (i) a vibrational bath coupled to the reaction coordinate,   
 $
 Q=x
 $; 
 (ii)  an electronic bath introducing energy gap fluctuations,   
 $
Q =\Pi_1 = \sum_{\bf n}  |{\bf R_{\bf n}}, \phi_{1} ({\bf r};{\bf R_{\bf n}})\ra \la \phi_{1	} ({\bf r};{\bf R_{\bf n}}), {\bf R_{\bf n}} |$ with $\Pi_1$ being the projection operator of the lower  adiabatic electronic state.
We employ the Drude bath spectral density 
\begin{align}
  J(\omega)=\frac{2\lambda\gamma\omega}{\omega^2+\gamma^2},  
  \end{align}
where $\lambda$ is the coupling strength,   $\gamma$ is the decay rate.
  For the electronic bath,   the solvent will reorganize when the electronic state changes character. The molecular Hamiltonian is modified by the reorganization of solvent
$
  H_{\tS}\rightarrow H_{\tS}+\lambda \sum_{\bf n}|{\bf R_{\bf n}}, \phi_1({\bf R_{\bf n}})\ra\la \phi_1( {\bf R_{\bf n}}), {\bf R_{\bf n}} | 
$
 for the Drude spectrum,   reducing the energy gap between ${\rm S}_1$ and ${\rm S}_2$. (Details in Appendix A)

\begin{figure*}[htbp]
  \centering
  \includegraphics[width=\textwidth]{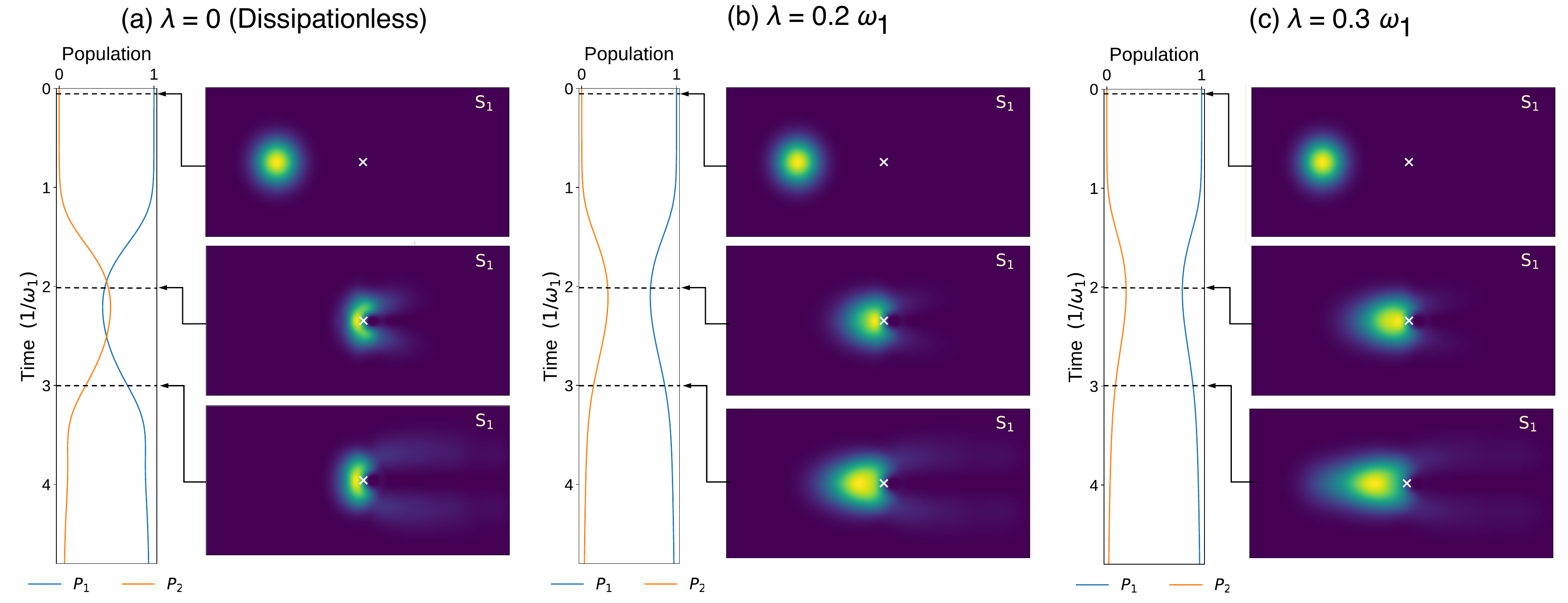}
  \caption{Nuclear wavepacket \& electron population dynamics with a vibrational bath  at different coupling strength $\lambda$.  $\omega_1^{-1}$ is chosen as the time unit,    parameters: $\gamma=1$,   $T=1$ in $\omega_1$. The CI is marked by $\times$.}
  \label{fig_xB}
\end{figure*}

\begin{figure*}[htbp]
  \centering
  \includegraphics[width=\textwidth]{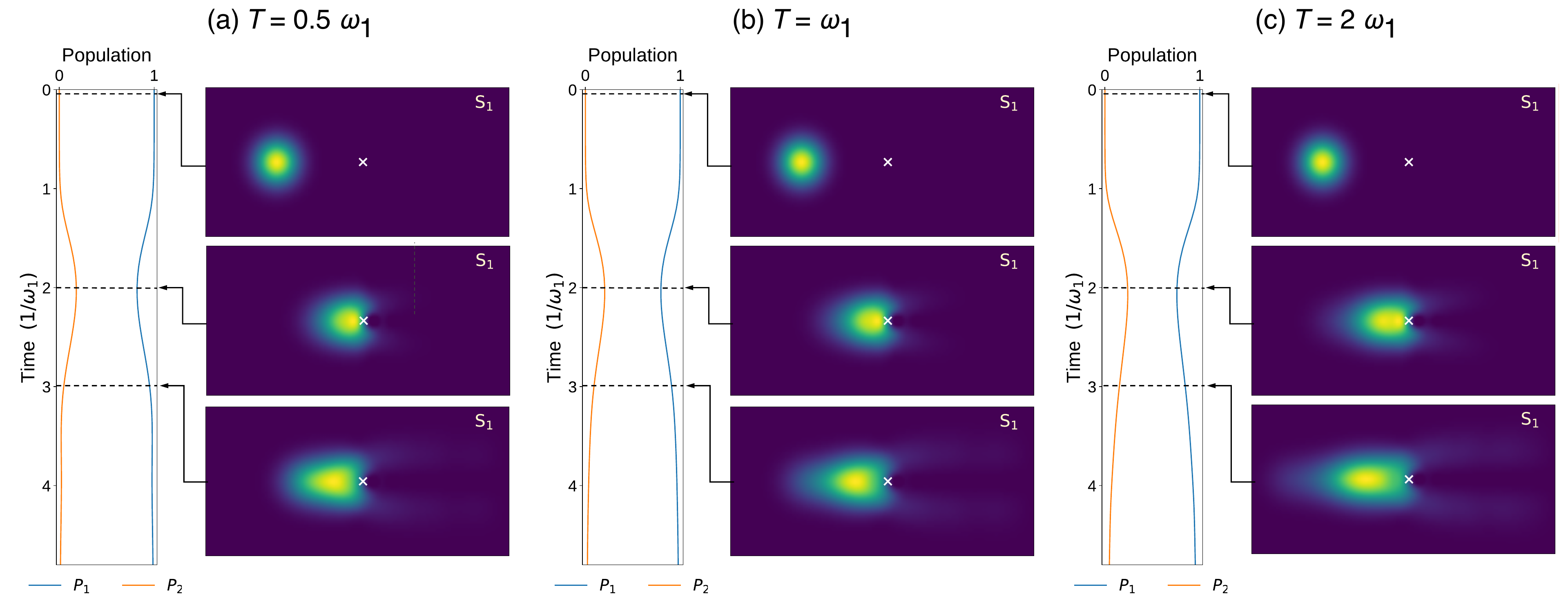}
  \caption{Nuclear wavepacket \& electron population dynamics with vibrational bath coupling at different temperature $T$. $\omega_1^{-1}$ is chosen as the time unit,     parameters: $\gamma=1$,   $\lambda=0.3$ in $\omega_1$. The CI is marked by $\times$.}
  \label{fig_xb_temp}
\end{figure*}

\begin{figure*}[htbp]
  \centering
  \includegraphics[width=\textwidth]{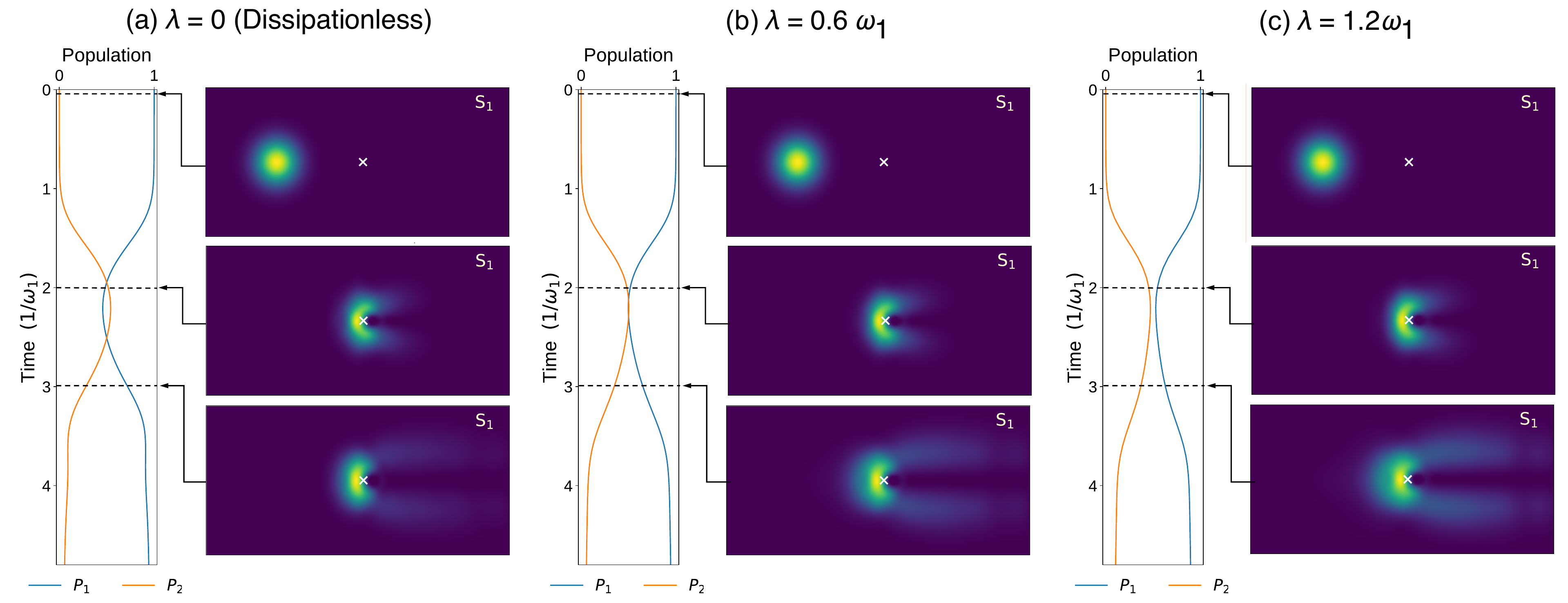}
  \caption{Nuclear wavepacket and electron population dynamics with electronic bath coupling at different coupling strength $\lambda$.  $\omega_1^{-1}$ is chosen as the time unit,    parameters: $\gamma=1$,   $T=1$ in $\omega_1$. The CI is marked by $\times$.}
  \label{fig_eB}
\end{figure*}

\begin{figure*}[htbp]
 \centering
 \includegraphics[width=\textwidth]{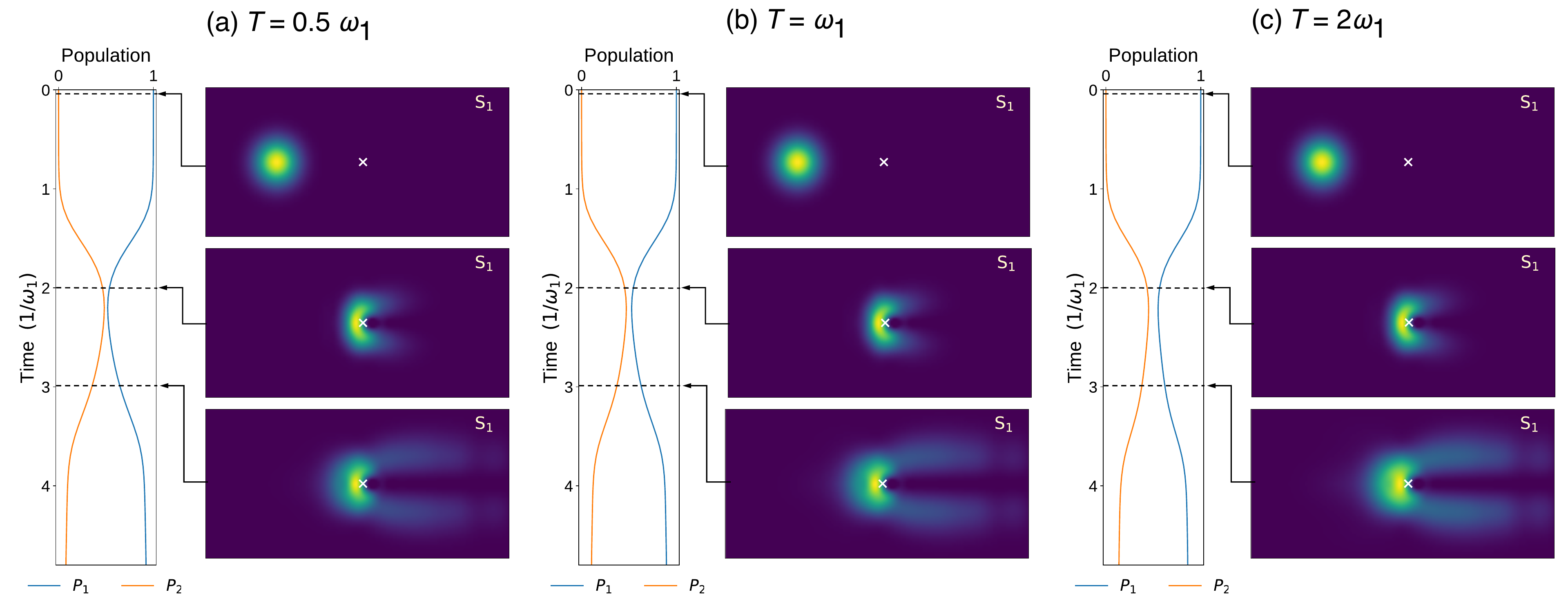}
 \caption{Nuclear wavepacket \& electron population dynamics with  electronic bath coupling at different temperature $T$.  $\omega_1^{-1}$ is chosen as the time unit,    parameters: $\gamma=1$,   $\lambda=1.2$ in $\omega_1$.  The CI is marked by $\times$.}
 \label{fig_eb_temp}
\end{figure*}

The initial state is  set as a Gaussian wavepacket,  
\be
\chi_0(x, y) =\frac{1}{\sqrt{\pi}}\exp{-\half \qty((x - x_0)^2+(y - y_0)^2)}e^{i p_x x}
\ee
centered at $(x_0, y_0) = (-4,  0)$ on the lower state ${\rm S}_1$  with initial momentum $p_x=2$. 
\Fig{fig_xB} depicts the electronic population and nuclear wave packet dynamics  $p_\alpha(\bf R, t) =  \rho_{\alpha \alpha}(\bf R, \bf R, t)$ with  vibrational relaxation at different system-bath coupling strength $\lambda$.  
Without the bath, as  the nuclear wavepacket  reaches the CI on ${\rm S}_1$,   there is  significant nonadiabatic transitions and a nodel line along $y=0$ in the probability distribution, which is a hallmark of the  geometric phase.
With the vibrational bath,  the diffusion of the wavepacket along $x$-direction is enhanced,   and  the nonadiabatic transitions is weakened. 
\Fig{fig_xb_temp} depicts the electron population and wavepacket dynamics with  vibrational bath coupling at different temperature $T$.  
As the temperature increase,   we can see the diffusion of the wavepacket along $x$-direction,   and  transition of electron is enhanced. However,   the nodal line in the interference pattern induced by the  geometric phase remains intact.

\Fig{fig_eB} depicts the electron population and nuclear wavepacket dynamics with  an electronic bath  at different coupling strength $\lambda$. 
 When electronic bath coupling $\lambda$ increasing,   the growth of  the excited electron population on ${\rm S}_2$ is suppressed.  The electron population gets redistributed due to the reorganization of the solvent which modifies  the gap between PESs. 
These changes can also be observed when temperature $T$ increases,   as \Fig{fig_eb_temp} shows.
  However,   as coupling strength $\lambda$ and temperature $T$ increase,   the characteristic interference pattern of  geometric phase not only  survives,   but is further enhanced.

 \begin{figure*}[htbp]
  \centering
  \includegraphics[width=\textwidth]{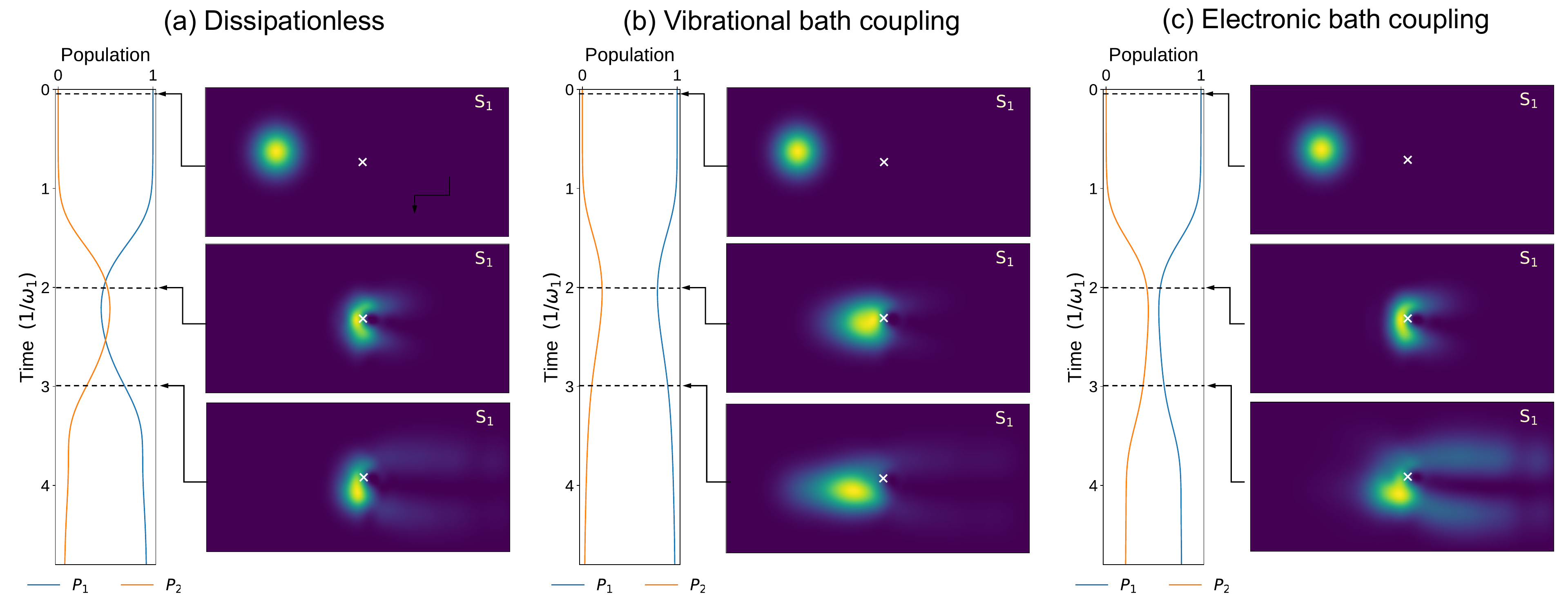}
  \caption{Nuclear wavepacket and electron population dynamics with an asymmetrical initial state. (a). dissipationless; (b) vibrational bath coupling with parameters: $\lambda=0.3$,    $\gamma=1$,   $T=1$ in $\omega_1$. (c) electronic bath coupling with parameters: $\lambda=2$,   $\gamma=1$,   $T=1$ in $\omega_1$.  $\omega_1^{-1}$ is chosen as the time unit. The CI is marked by $\times$.}
  \label{fig_ns}
\end{figure*}

We choose an asymmetrical initial nuclear wavepacket by shifting the center of the Gaussian wavepacket to $(-4,  0.5)$. 
 The interference pattern is still clearly visible despite strong dissipation for both electronic  and vibrational baths (\Fig{fig_ns}).  We have used a stronger  molecule-solvent interaction  than typical conditions.
 This indicates that the robustness of the interference pattern is not due to symmetry.

The robustness of the geometric phase-induced quantum interference can be understood using the path integral picture in Sec. \ref{sec:theory}.
Consider the two-dimensional system with $y$-reflection symmetry as in \Fig{fig_path}.   For   APES $V(x,  y)$ with  $V(x,  y) = V(x,  -y)$,   when adiabatic path pair $\Gamma_+$ and $\Gamma_-$ surrounding CI and  connecting ${\bf R}_0$ and ${\bf R}_f$ are symmetric about $y=0$,   their dynamical actions are  equal ${S}_{}[\Gamma_+(t)]={S}_{}[\Gamma_-(t)]$. With the phase difference between $\Gamma_+$ and $\Gamma_-$,    $W[\Gamma_+]+W[\Gamma_-]=W[\Gamma_-](1+e^{i\pi})=0$, as this applies to all pairs of paths,    there will always be a destructive interference leading to a nodal line in the nuclear distribution along $y=0$ even in the presence of dissipation.


For the vibrational bath  $Q=x$,   the influence functionals of the  path pair $\Gamma_+$ and $\Gamma_-$ are equal 
\begin{align}
\mathcal{F}[x^+(t),  x'(t)]=\mathcal{F}[x^-(t),  x'(t)] 
\end{align}
 the propagator connecting ${\bf R}_0$ and ${\bf R}_f$ can be presented as a sum of all the path pairs $\{\Gamma_+,  \Gamma_-\}$
\begin{align}
&\mathcal{G}_{}({\bf R}_{f},  {\rm S}_1,  {\bf R}_{f},  {\rm S}_1,  t;{\bf R}_{0},  {\rm S}_1,  {\bf R}_{0},  {\rm S}_1,  t_0)=
\nl
&\sum_{\{{\bf R}^{\pm},  {\bf R}_j'\}} \mathcal{F}_{\pm}[x_j,  x'_j] \sum_{\{ \alpha_j,    \alpha_j'\}}^{{\rm S}_1\rightarrow {\rm S}_1}
\qty{ W[\xi^-(t)]e^{iS[\xi^-(t)]} 
+W[\xi^+(t) ]e^{iS[\xi^+(t) ]} }
 \overline{W} [\xi'(t)]e^{-iS[\xi'(t)]}  = 0
\end{align}
As the influence functional of the two paths are equal, it can be factored out such that the transition amplitudes interfere destructively. 

The same argument can be applied to the electronic bath  $Q = \Pi_1$,   as the influence functional remain equal for the pair of paths with reflection symmetry. 
%
Therefore, we have demonstrated that,  both the vibrational and  electronic bath do  not eliminate the destructive interference of different path pairs $\{\Gamma_- ,   \Gamma_+\}$. 
Thus as the dissipation strengthening,   the  geometric phase-induced quantum interference can remain robust.
Although this analysis is only rigorously valid in the close vicinity of a conical intersection, it provides useful insights to understand the computational results. 

\section{Conclusion}
 Through numerically exact dissipative conical intersection dynamics modeling, we have 
 demonstrated that quantum decoherence does not eliminate the quantum interference originated from the  geometric phase. Although the population dynamics can be influenced by the bath, the geometric phase effects is highly robust against both vibrational relaxation and electronic dephasing. 
 Further simulations with asymmetric initial  conditions show that the interference pattern surviving from the dissipation is protected by the topology of the CI,  not only by the space reflex symmetry.  

Our results  suggest that geometric phase effects, neglected in most mixed quantum-classical methods may be important even in condensed phase environments. 

\section{Acknowledgement}
This work is supported by the National Natural Science
Foundation of China (Grant Nos. 22473090 and
92356310).
We appreciate the help and advice from Y. Su,   Y. Wang,   Y.J. Xie,   L.Z. Ye and X.T. Zhu.


\clearpage
\appendix

\section{Details about the electronic bath coupling}

The total Hamiltonian of  bath-electronic state coupling reads
\begin{align} \label{heb}
H_{\rm e-\B}=\sum_{{\bf n},j}&\{\frac{1}{2}\omega^2_j (x_j+d_j({\bf R_{\bf n}}))^2 | {\bf R_{\bf n}},\phi_1({\bf R_{\bf n}})\ra\la \phi_1 ({\bf R_{\bf n}}),{\bf R_{\bf n}} | 
+\frac{1}{2} \omega^2_j  x_j^2 | {\bf R_{\bf n}},\phi_2({\bf R_{\bf n}})\ra\la \phi_2( {\bf R_{\bf n}}), {\bf R_{\bf n}}|\}
\nl
=\sum_{{\bf n},j}&\{ \frac{1}{2} \omega^2_j  x^2_j \otimes(|{\bf R_{\bf n}}, \phi_1({\bf R_{\bf n}})\ra\la \phi_1( {\bf R_{\bf n}}),{\bf R_{\bf n}} | + | {\bf R_{\bf n}},\phi_2({\bf R_{\bf n}})\ra\la \phi_2 ({\bf R_{\bf n}}),{\bf R_{\bf n}} |)
\nl
&+ \omega^2_j d_j({\bf R_{\bf n}})  x_j  |{\bf R_{\bf n}}, \phi_1({\bf R_{\bf n}})\ra\la \phi_1 ({\bf R_{\bf n}}),{\bf R_{\bf n}} | 
+\frac{1}{2} \omega^2_j  d^2_j({\bf R_{\bf n}}) |{\bf R_{\bf n}}, \phi_1({\bf R_{\bf n}})\ra\la \phi_1( {\bf R_{\bf n}}), {\bf R_{\bf n}} |\}.
\end{align}
Here $\{x_j \}$ represent the microscopic degrees of freedoms of the bosonic bath and $\{ d_j({\bf R_{\bf n}}) \}$ represent the displacement of solvent.

For simplification,  we make the approximation that the displacement of solvent $d_j({\bf R_{\bf n}})$ is independent of the nuclear configuration $d_j({\bf R_{\bf n}})\approx d_j$.   
 Under the approximation, the terms in \Eq{heb} can be rearranged as
\begin{align}
\sum_{{\bf n},j} & \frac{1}{2} \omega^2_j  x^2_j \otimes(|{\bf R_{\bf n}}, \phi_1({\bf R_{\bf n}})\ra\la \phi_1( {\bf R_{\bf n}}), {\bf R_{\bf n}} | + |{\bf R_{\bf n}}, \phi_2({\bf R_{\bf n}})\ra\la \phi_2 ({\bf R_{\bf n}}),{\bf R_{\bf n}} |)
\nl
=\sum_{j} & \frac{1}{2} \omega^2_j  x^2_j \otimes\sum_{{\bf R_{\bf n}}} \big( |{\bf R_{\bf n}}, \phi_1({\bf R_{\bf n}})\ra\la \phi_1 ({\bf R_{\bf n}}), {\bf n}| + |{\bf R_{\bf n}}, \phi_2({\bf R_{\bf n}})\ra\la \phi_2 ({\bf R_{\bf n}}), {\bf R_{\bf n}}| \big) = H_{\B} \otimes I_{\M},
\end{align}
\begin{align}
&\sum_{{\bf n},j}
  \omega^2_j d_j({\bf R_{\bf n}})  x_j  |{\bf R_{\bf n}}, \phi_1({\bf R_{\bf n}})\ra\la \phi_1( {\bf R_{\bf n}}), {\bf R_{\bf n}} | 
\nl  
  \approx& \sum_{j}  \omega^2_j d_j  x_j \otimes \sum_{\bf n} |{\bf R_{\bf n}}, \phi_1({\bf R_{\bf n}})\ra\la \phi_1( {\bf R_{\bf n}}), {\bf R_{\bf n}} |  = X\otimes  \sum_{\bf n}|{\bf R_{\bf n}}, \phi_1({\bf R_{\bf n}})\ra\la \phi_1( {\bf R_{\bf n}}), {\bf R_{\bf n}} | ,
\end{align}
and
\begin{align}
&\sum_{{\bf n},j}
\frac{1}{2} \omega^2_j  d^2_j({\bf R_{\bf n}})|{\bf R_{\bf n}}, \phi_1({\bf R_{\bf n}})\ra\la \phi_1( {\bf R_{\bf n}}), {\bf R_{\bf n}} |  
\nl
\approx &\sum_{j} \frac{1}{2} \omega^2_j  d^2_j \sum_{{\bf n}} |{\bf R_{\bf n}}, \phi_1({\bf R_{\bf n}})\ra\la \phi_1( {\bf R_{\bf n}}), {\bf R_{\bf n}} |  = \lambda  \sum_{\bf n}|{\bf R_{\bf n}}, \phi_1({\bf R_{\bf n}})\ra\la \phi_1( {\bf R_{\bf n}}), {\bf R_{\bf n}} |  ,
\end{align}
for Drude spectrum.

Thus the total Hamiltonian of  bath-electronic state coupling can be decomposed as 
\begin{align}
H_{\rm e-\B} = &H_{\B} +  X\otimes  \sum_{\bf n}|{\bf R_{\bf n}}, \phi_1({\bf R_{\bf n}})\ra\la \phi_1( {\bf R_{\bf n}}), {\bf R_{\bf n}} |  +\lambda  \sum_{\bf n}|{\bf R_{\bf n}}, \phi_1({\bf R_{\bf n}})\ra\la \phi_1( {\bf R_{\bf n}}), {\bf R_{\bf n}} |  .
\end{align}

\bibliography{bibrefs,ALDR,../dynamics,topology, OQS, qchem,../cavity,../topology,../qchem}

\begin{thebibliography}{60}%
\makeatletter
\providecommand \@ifxundefined [1]{%
 \@ifx{#1\undefined}
}%
\providecommand \@ifnum [1]{%
 \ifnum #1\expandafter \@firstoftwo
 \else \expandafter \@secondoftwo
 \fi
}%
\providecommand \@ifx [1]{%
 \ifx #1\expandafter \@firstoftwo
 \else \expandafter \@secondoftwo
 \fi
}%
\providecommand \natexlab [1]{#1}%
\providecommand \enquote  [1]{``#1''}%
\providecommand \bibnamefont  [1]{#1}%
\providecommand \bibfnamefont [1]{#1}%
\providecommand \citenamefont [1]{#1}%
\providecommand \href@noop [0]{\@secondoftwo}%
\providecommand \href [0]{\begingroup \@sanitize@url \@href}%
\providecommand \@href[1]{\@@startlink{#1}\@@href}%
\providecommand \@@href[1]{\endgroup#1\@@endlink}%
\providecommand \@sanitize@url [0]{\catcode `\\12\catcode `\$12\catcode
  `\&12\catcode `\#12\catcode `\^12\catcode `\_12\catcode `\%12\relax}%
\providecommand \@@startlink[1]{}%
\providecommand \@@endlink[0]{}%
\providecommand \url  [0]{\begingroup\@sanitize@url \@url }%
\providecommand \@url [1]{\endgroup\@href {#1}{\urlprefix }}%
\providecommand \urlprefix  [0]{URL }%
\providecommand \Eprint [0]{\href }%
\providecommand \doibase [0]{https://doi.org/}%
\providecommand \selectlanguage [0]{\@gobble}%
\providecommand \bibinfo  [0]{\@secondoftwo}%
\providecommand \bibfield  [0]{\@secondoftwo}%
\providecommand \translation [1]{[#1]}%
\providecommand \BibitemOpen [0]{}%
\providecommand \bibitemStop [0]{}%
\providecommand \bibitemNoStop [0]{.\EOS\space}%
\providecommand \EOS [0]{\spacefactor3000\relax}%
\providecommand \BibitemShut  [1]{\csname bibitem#1\endcsname}%
\let\auto@bib@innerbib\@empty
\bibitem [{\citenamefont {Domcke}\ and\ \citenamefont
  {Yarkony}(2012)}]{domcke2012}%
  \BibitemOpen
  \bibfield  {author} {\bibinfo {author} {\bibfnamefont {W.}~\bibnamefont
  {Domcke}}\ and\ \bibinfo {author} {\bibfnamefont {D.~R.}\ \bibnamefont
  {Yarkony}},\ }\bibfield  {title} {\enquote {\bibinfo {title} {Role of
  {{Conical Intersections}} in {{Molecular Spectroscopy}} and {{Photoinduced
  Chemical Dynamics}}},}\ }\href
  {https://doi.org/10.1146/annurev-physchem-032210-103522} {\bibfield
  {journal} {\bibinfo  {journal} {Annu. Rev. Phys. Chem.}\ }\textbf {\bibinfo
  {volume} {63}},\ \bibinfo {pages} {325--352} (\bibinfo {year}
  {2012})}\BibitemShut {NoStop}%
\bibitem [{\citenamefont {K{\"o}ppel}, \citenamefont {Domcke},\ and\
  \citenamefont {Cederbaum}(2004)}]{koppel2004}%
  \BibitemOpen
  \bibfield  {author} {\bibinfo {author} {\bibfnamefont {H.}~\bibnamefont
  {K{\"o}ppel}}, \bibinfo {author} {\bibfnamefont {W.}~\bibnamefont {Domcke}},\
  and\ \bibinfo {author} {\bibfnamefont {L.~S.}\ \bibnamefont {Cederbaum}},\
  }\enquote {\bibinfo {title} {{{THE MULTI-MODE VIBRONIC-COUPLING
  APPROACH}}},}\ in\ \href {https://doi.org/10.1142/9789812565464_0007} {\emph
  {\bibinfo {booktitle} {Advanced {{Series}} in {{Physical Chemistry}}}}},\
  Vol.~\bibinfo {volume} {15}\ (\bibinfo  {publisher} {WORLD SCIENTIFIC},\
  \bibinfo {year} {2004})\ pp.\ \bibinfo {pages} {323--367}\BibitemShut
  {NoStop}%
\bibitem [{\citenamefont {Larson}, \citenamefont {Sj{\"o}qvist},\ and\
  \citenamefont {{\"O}hberg}(2020)}]{larson2020}%
  \BibitemOpen
  \bibfield  {author} {\bibinfo {author} {\bibfnamefont {J.}~\bibnamefont
  {Larson}}, \bibinfo {author} {\bibfnamefont {E.}~\bibnamefont
  {Sj{\"o}qvist}},\ and\ \bibinfo {author} {\bibfnamefont {P.}~\bibnamefont
  {{\"O}hberg}},\ }\href {https://doi.org/10.1007/978-3-030-34882-3} {\emph
  {\bibinfo {title} {Conical {{Intersections}} in {{Physics}}: {{An
  Introduction}} to {{Synthetic Gauge Theories}}}}},\ \bibinfo {series}
  {Lecture {{Notes}} in {{Physics}}}, Vol.\ \bibinfo {volume} {965}\ (\bibinfo
  {publisher} {Springer International Publishing},\ \bibinfo {address} {Cham},\
  \bibinfo {year} {2020})\BibitemShut {NoStop}%
\bibitem [{\citenamefont {Levine}\ and\ \citenamefont
  {Mart{\'i}nez}(2007)}]{levine2007}%
  \BibitemOpen
  \bibfield  {author} {\bibinfo {author} {\bibfnamefont {B.~G.}\ \bibnamefont
  {Levine}}\ and\ \bibinfo {author} {\bibfnamefont {T.~J.}\ \bibnamefont
  {Mart{\'i}nez}},\ }\bibfield  {title} {\enquote {\bibinfo {title}
  {Isomerization {{Through Conical Intersections}}},}\ }\href
  {https://doi.org/10.1146/annurev.physchem.57.032905.104612} {\bibfield
  {journal} {\bibinfo  {journal} {Annu. Rev. Phys. Chem.}\ }\textbf {\bibinfo
  {volume} {58}},\ \bibinfo {pages} {613--634} (\bibinfo {year}
  {2007})}\BibitemShut {NoStop}%
\bibitem [{\citenamefont {Mead}\ and\ \citenamefont
  {Truhlar}(1979)}]{mead1979}%
  \BibitemOpen
  \bibfield  {author} {\bibinfo {author} {\bibfnamefont {C.~A.}\ \bibnamefont
  {Mead}}\ and\ \bibinfo {author} {\bibfnamefont {D.~G.}\ \bibnamefont
  {Truhlar}},\ }\bibfield  {title} {\enquote {\bibinfo {title} {On the
  determination of {{Born}}--{{Oppenheimer}} nuclear motion wave functions
  including complications due to conical intersections and identical nuclei},}\
  }\href {https://doi.org/10.1063/1.437734} {\bibfield  {journal} {\bibinfo
  {journal} {The Journal of Chemical Physics}\ }\textbf {\bibinfo {volume}
  {70}},\ \bibinfo {pages} {2284--2296} (\bibinfo {year} {1979})}\BibitemShut
  {NoStop}%
\bibitem [{\citenamefont {Mead}(1992{\natexlab{a}})}]{mead1992}%
  \BibitemOpen
  \bibfield  {author} {\bibinfo {author} {\bibfnamefont {C.~A.}\ \bibnamefont
  {Mead}},\ }\bibfield  {title} {\enquote {\bibinfo {title} {The geometric
  phase in molecular systems},}\ }\href
  {https://doi.org/10.1103/RevModPhys.64.51} {\bibfield  {journal} {\bibinfo
  {journal} {Rev. Mod. Phys.}\ }\textbf {\bibinfo {volume} {64}},\ \bibinfo
  {pages} {51--85} (\bibinfo {year} {1992}{\natexlab{a}})}\BibitemShut
  {NoStop}%
\bibitem [{\citenamefont {{Joubert-Doriol}}, \citenamefont {Ryabinkin},\ and\
  \citenamefont {Izmaylov}(2013)}]{joubert-doriol2013}%
  \BibitemOpen
  \bibfield  {author} {\bibinfo {author} {\bibfnamefont {L.}~\bibnamefont
  {{Joubert-Doriol}}}, \bibinfo {author} {\bibfnamefont {I.~G.}\ \bibnamefont
  {Ryabinkin}},\ and\ \bibinfo {author} {\bibfnamefont {A.~F.}\ \bibnamefont
  {Izmaylov}},\ }\bibfield  {title} {\enquote {\bibinfo {title} {Geometric
  phase effects in low-energy dynamics near conical intersections: {{A}} study
  of the multidimensional linear vibronic coupling model},}\ }\href
  {https://doi.org/10.1063/1.4844095} {\bibfield  {journal} {\bibinfo
  {journal} {J. Chem. Phys.}\ }\textbf {\bibinfo {volume} {139}},\ \bibinfo
  {pages} {234103} (\bibinfo {year} {2013})}\BibitemShut {NoStop}%
\bibitem [{\citenamefont {Bersuker}(2006)}]{bersuker2006}%
  \BibitemOpen
  \bibfield  {author} {\bibinfo {author} {\bibfnamefont {I.}~\bibnamefont
  {Bersuker}},\ }\href {https://doi.org/10.1017/CBO9780511524769} {\emph
  {\bibinfo {title} {The {{Jahn-Teller Effect}}}}}\ (\bibinfo  {publisher}
  {Cambridge University Press},\ \bibinfo {address} {Cambridge},\ \bibinfo
  {year} {2006})\BibitemShut {NoStop}%
\bibitem [{\citenamefont {Gu}\ and\ \citenamefont
  {Mukamel}(2020{\natexlab{a}})}]{gu2020b}%
  \BibitemOpen
  \bibfield  {author} {\bibinfo {author} {\bibfnamefont {B.}~\bibnamefont
  {Gu}}\ and\ \bibinfo {author} {\bibfnamefont {S.}~\bibnamefont {Mukamel}},\
  }\bibfield  {title} {\enquote {\bibinfo {title} {Cooperative {{Conical
  Intersection Dynamics}} of {{Two Pyrazine Molecules}} in an {{Optical
  Cavity}}},}\ }\href {https://doi.org/10.1021/acs.jpclett.0c00381} {\bibfield
  {journal} {\bibinfo  {journal} {J. Phys. Chem. Lett.}\ }\textbf {\bibinfo
  {volume} {11}},\ \bibinfo {pages} {5555--5562} (\bibinfo {year}
  {2020}{\natexlab{a}})}\BibitemShut {NoStop}%
\bibitem [{\citenamefont {Gu}\ and\ \citenamefont
  {Mukamel}(2020{\natexlab{b}})}]{gu2020c}%
  \BibitemOpen
  \bibfield  {author} {\bibinfo {author} {\bibfnamefont {B.}~\bibnamefont
  {Gu}}\ and\ \bibinfo {author} {\bibfnamefont {S.}~\bibnamefont {Mukamel}},\
  }\bibfield  {title} {\enquote {\bibinfo {title} {Manipulating nonadiabatic
  conical intersection dynamics by optical cavities},}\ }\href
  {https://doi.org/10.1039/C9SC04992D} {\bibfield  {journal} {\bibinfo
  {journal} {Chem. Sci.}\ }\textbf {\bibinfo {volume} {11}},\ \bibinfo {pages}
  {1290--1298} (\bibinfo {year} {2020}{\natexlab{b}})}\BibitemShut {NoStop}%
\bibitem [{\citenamefont {Domcke}, \citenamefont {Yarkony},\ and\ \citenamefont
  {K{\"o}ppel}(2011)}]{domcke2011}%
  \BibitemOpen
  \bibfield  {author} {\bibinfo {author} {\bibfnamefont {W.}~\bibnamefont
  {Domcke}}, \bibinfo {author} {\bibfnamefont {D.~R.}\ \bibnamefont
  {Yarkony}},\ and\ \bibinfo {author} {\bibfnamefont {H.}~\bibnamefont
  {K{\"o}ppel}},\ }\href@noop {} {\emph {\bibinfo {title} {Conical
  {{Intersections}}: {{Theory}}, {{Computation}} and {{Experiment}}}}}\
  (\bibinfo  {publisher} {World Scientific},\ \bibinfo {year}
  {2011})\BibitemShut {NoStop}%
\bibitem [{\citenamefont {Prigogine}\ and\ \citenamefont
  {Rice}(2002)}]{prigogine2002}%
  \BibitemOpen
  \bibfield  {author} {\bibinfo {author} {\bibfnamefont {I.}~\bibnamefont
  {Prigogine}}\ and\ \bibinfo {author} {\bibfnamefont {S.~A.}\ \bibnamefont
  {Rice}},\ }\href@noop {} {\emph {\bibinfo {title} {The {{Role}} of
  {{Degenerate States}} in {{Chemistry}}, {{Volume}} 124}}},\ \bibinfo
  {edition} {1st}\ ed.,\ edited by\ \bibinfo {editor} {\bibfnamefont
  {M.}~\bibnamefont {Baer}}\ and\ \bibinfo {editor} {\bibfnamefont {G.~D.}\
  \bibnamefont {Billing}}\ (\bibinfo  {publisher} {Wiley-Interscience},\
  \bibinfo {address} {Hoboken, N.J},\ \bibinfo {year} {2002})\BibitemShut
  {NoStop}%
\bibitem [{\citenamefont {Mead}\ and\ \citenamefont
  {Truhlar}(1982)}]{mead1982}%
  \BibitemOpen
  \bibfield  {author} {\bibinfo {author} {\bibfnamefont {C.~A.}\ \bibnamefont
  {Mead}}\ and\ \bibinfo {author} {\bibfnamefont {D.~G.}\ \bibnamefont
  {Truhlar}},\ }\bibfield  {title} {\enquote {\bibinfo {title} {Conditions for
  the definition of a strictly diabatic electronic basis for molecular
  systems},}\ }\href {https://doi.org/10.1063/1.443853} {\bibfield  {journal}
  {\bibinfo  {journal} {The Journal of Chemical Physics}\ }\textbf {\bibinfo
  {volume} {77}},\ \bibinfo {pages} {6090--6098} (\bibinfo {year}
  {1982})}\BibitemShut {NoStop}%
\bibitem [{\citenamefont {Casida}, \citenamefont {Natarajan},\ and\
  \citenamefont {Deutsch}(2012)}]{casida2012}%
  \BibitemOpen
  \bibfield  {author} {\bibinfo {author} {\bibfnamefont {M.~E.}\ \bibnamefont
  {Casida}}, \bibinfo {author} {\bibfnamefont {B.}~\bibnamefont {Natarajan}},\
  and\ \bibinfo {author} {\bibfnamefont {T.}~\bibnamefont {Deutsch}},\
  }\bibfield  {title} {\enquote {\bibinfo {title} {Non-{{Born}}--{{Oppenheimer
  Dynamics}} and {{Conical Intersections}}},}\ }in\ \href
  {https://doi.org/10.1007/978-3-642-23518-4_14} {\emph {\bibinfo {booktitle}
  {Fundamentals of {{Time-Dependent Density Functional Theory}}}}},\ \bibinfo
  {series and number} {Lecture {{Notes}} in {{Physics}}},\ \bibinfo {editor}
  {edited by\ \bibinfo {editor} {\bibfnamefont {M.~A.}\ \bibnamefont
  {Marques}}, \bibinfo {editor} {\bibfnamefont {N.~T.}\ \bibnamefont {Maitra}},
  \bibinfo {editor} {\bibfnamefont {F.~M.}\ \bibnamefont {Nogueira}}, \bibinfo
  {editor} {\bibfnamefont {E.}~\bibnamefont {Gross}},\ and\ \bibinfo {editor}
  {\bibfnamefont {A.}~\bibnamefont {Rubio}}}\ (\bibinfo  {publisher}
  {Springer},\ \bibinfo {address} {Berlin, Heidelberg},\ \bibinfo {year}
  {2012})\ pp.\ \bibinfo {pages} {279--299}\BibitemShut {NoStop}%
\bibitem [{\citenamefont {Polli}\ \emph {et~al.}(2010)\citenamefont {Polli},
  \citenamefont {Alto{\`e}}, \citenamefont {Weingart}, \citenamefont
  {Spillane}, \citenamefont {Manzoni}, \citenamefont {Brida}, \citenamefont
  {Tomasello}, \citenamefont {Orlandi}, \citenamefont {Kukura}, \citenamefont
  {Mathies}, \citenamefont {Garavelli},\ and\ \citenamefont
  {Cerullo}}]{polli2010}%
  \BibitemOpen
  \bibfield  {author} {\bibinfo {author} {\bibfnamefont {D.}~\bibnamefont
  {Polli}}, \bibinfo {author} {\bibfnamefont {P.}~\bibnamefont {Alto{\`e}}},
  \bibinfo {author} {\bibfnamefont {O.}~\bibnamefont {Weingart}}, \bibinfo
  {author} {\bibfnamefont {K.~M.}\ \bibnamefont {Spillane}}, \bibinfo {author}
  {\bibfnamefont {C.}~\bibnamefont {Manzoni}}, \bibinfo {author} {\bibfnamefont
  {D.}~\bibnamefont {Brida}}, \bibinfo {author} {\bibfnamefont
  {G.}~\bibnamefont {Tomasello}}, \bibinfo {author} {\bibfnamefont
  {G.}~\bibnamefont {Orlandi}}, \bibinfo {author} {\bibfnamefont
  {P.}~\bibnamefont {Kukura}}, \bibinfo {author} {\bibfnamefont {R.~A.}\
  \bibnamefont {Mathies}}, \bibinfo {author} {\bibfnamefont {M.}~\bibnamefont
  {Garavelli}},\ and\ \bibinfo {author} {\bibfnamefont {G.}~\bibnamefont
  {Cerullo}},\ }\bibfield  {title} {\enquote {\bibinfo {title} {Conical
  intersection dynamics of the primary photoisomerization event in vision},}\
  }\href {https://doi.org/10.1038/nature09346} {\bibfield  {journal} {\bibinfo
  {journal} {Nature}\ }\textbf {\bibinfo {volume} {467}},\ \bibinfo {pages}
  {440--443} (\bibinfo {year} {2010})}\BibitemShut {NoStop}%
\bibitem [{\citenamefont {Mandal}\ \emph {et~al.}(2023)\citenamefont {Mandal},
  \citenamefont {Taylor}, \citenamefont {Weight}, \citenamefont {Koessler},
  \citenamefont {Li},\ and\ \citenamefont {Huo}}]{mandal2023}%
  \BibitemOpen
  \bibfield  {author} {\bibinfo {author} {\bibfnamefont {A.}~\bibnamefont
  {Mandal}}, \bibinfo {author} {\bibfnamefont {M.~A.}\ \bibnamefont {Taylor}},
  \bibinfo {author} {\bibfnamefont {B.~M.}\ \bibnamefont {Weight}}, \bibinfo
  {author} {\bibfnamefont {E.~R.}\ \bibnamefont {Koessler}}, \bibinfo {author}
  {\bibfnamefont {X.}~\bibnamefont {Li}},\ and\ \bibinfo {author}
  {\bibfnamefont {P.}~\bibnamefont {Huo}},\ }\bibfield  {title} {\enquote
  {\bibinfo {title} {Theoretical {{Advances}} in {{Polariton Chemistry}} and
  {{Molecular Cavity Quantum Electrodynamics}}},}\ }\href
  {https://doi.org/10.1021/acs.chemrev.2c00855} {\bibfield  {journal} {\bibinfo
   {journal} {Chem. Rev.}\ }\textbf {\bibinfo {volume} {123}},\ \bibinfo
  {pages} {9786--9879} (\bibinfo {year} {2023})}\BibitemShut {NoStop}%
\bibitem [{\citenamefont {Gu}\ and\ \citenamefont
  {Mukamel}(2020{\natexlab{c}})}]{gu2020a}%
  \BibitemOpen
  \bibfield  {author} {\bibinfo {author} {\bibfnamefont {B.}~\bibnamefont
  {Gu}}\ and\ \bibinfo {author} {\bibfnamefont {S.}~\bibnamefont {Mukamel}},\
  }\bibfield  {title} {\enquote {\bibinfo {title} {Manipulating
  {{Two-Photon-Absorption}} of {{Cavity Polaritons}} by {{Entangled Light}}},}\
  }\href {https://doi.org/10.1021/acs.jpclett.0c02282} {\bibfield  {journal}
  {\bibinfo  {journal} {J. Phys. Chem. Lett.}\ }\textbf {\bibinfo {volume}
  {11}},\ \bibinfo {pages} {8177--8182} (\bibinfo {year}
  {2020}{\natexlab{c}})}\BibitemShut {NoStop}%
\bibitem [{\citenamefont {Rafiq}\ \emph {et~al.}(2023)\citenamefont {Rafiq},
  \citenamefont {Weingartz}, \citenamefont {Kromer}, \citenamefont
  {Castellano},\ and\ \citenamefont {Chen}}]{rafiq2023}%
  \BibitemOpen
  \bibfield  {author} {\bibinfo {author} {\bibfnamefont {S.}~\bibnamefont
  {Rafiq}}, \bibinfo {author} {\bibfnamefont {N.~P.}\ \bibnamefont
  {Weingartz}}, \bibinfo {author} {\bibfnamefont {S.}~\bibnamefont {Kromer}},
  \bibinfo {author} {\bibfnamefont {F.~N.}\ \bibnamefont {Castellano}},\ and\
  \bibinfo {author} {\bibfnamefont {L.~X.}\ \bibnamefont {Chen}},\ }\bibfield
  {title} {\enquote {\bibinfo {title} {Spin--vibronic coherence drives
  singlet--triplet conversion},}\ }\href
  {https://doi.org/10.1038/s41586-023-06233-y} {\bibfield  {journal} {\bibinfo
  {journal} {Nature}\ }\textbf {\bibinfo {volume} {620}},\ \bibinfo {pages}
  {776--781} (\bibinfo {year} {2023})}\BibitemShut {NoStop}%
\bibitem [{\citenamefont {Quenneville}\ and\ \citenamefont
  {Mart{\'i}nez}(2003)}]{quenneville2003}%
  \BibitemOpen
  \bibfield  {author} {\bibinfo {author} {\bibfnamefont {J.}~\bibnamefont
  {Quenneville}}\ and\ \bibinfo {author} {\bibfnamefont {T.~J.}\ \bibnamefont
  {Mart{\'i}nez}},\ }\bibfield  {title} {\enquote {\bibinfo {title} {Ab
  {{Initio Study}} of {{Cis}}-{{Trans Photoisomerization}} in {{Stilbene}} and
  {{Ethylene}}},}\ }\href {https://doi.org/10.1021/jp021210w} {\bibfield
  {journal} {\bibinfo  {journal} {J. Phys. Chem. A}\ }\textbf {\bibinfo
  {volume} {107}},\ \bibinfo {pages} {829--837} (\bibinfo {year}
  {2003})}\BibitemShut {NoStop}%
\bibitem [{\citenamefont {Berry}\ and\ \citenamefont
  {Balazs}(1979)}]{berry1979}%
  \BibitemOpen
  \bibfield  {author} {\bibinfo {author} {\bibfnamefont {M.~V.}\ \bibnamefont
  {Berry}}\ and\ \bibinfo {author} {\bibfnamefont {N.~L.}\ \bibnamefont
  {Balazs}},\ }\bibfield  {title} {\enquote {\bibinfo {title} {Nonspreading
  wave packets},}\ }\href {https://doi.org/10.1119/1.11855} {\bibfield
  {journal} {\bibinfo  {journal} {Am. J. Phys.}\ }\textbf {\bibinfo {volume}
  {47}},\ \bibinfo {pages} {264--267} (\bibinfo {year} {1979})}\BibitemShut
  {NoStop}%
\bibitem [{\citenamefont {Berry}(1984)}]{berry1984}%
  \BibitemOpen
  \bibfield  {author} {\bibinfo {author} {\bibfnamefont {M.~V.}\ \bibnamefont
  {Berry}},\ }\bibfield  {title} {\enquote {\bibinfo {title} {Quantal phase
  factors accompanying adiabatic changes},}\ }\href
  {https://doi.org/10.1098/rspa.1984.0023} {\bibfield  {journal} {\bibinfo
  {journal} {Proceedings of the Royal Society of London. A. Mathematical and
  Physical Sciences}\ }\textbf {\bibinfo {volume} {392}},\ \bibinfo {pages}
  {45--57} (\bibinfo {year} {1984})}\BibitemShut {NoStop}%
\bibitem [{\citenamefont {Xie}, \citenamefont {Liu},\ and\ \citenamefont
  {Gu}(2025)}]{xie2025.11124}%
  \BibitemOpen
  \bibfield  {author} {\bibinfo {author} {\bibfnamefont {Y.}~\bibnamefont
  {Xie}}, \bibinfo {author} {\bibfnamefont {R.}~\bibnamefont {Liu}},\ and\
  \bibinfo {author} {\bibfnamefont {B.}~\bibnamefont {Gu}},\ }\href
  {https://arxiv.org/abs/2505.11124} {\enquote {\bibinfo {title} {Topological
  quantum molecular dynamics},}\ } (\bibinfo {year} {2025}),\ \Eprint
  {https://arxiv.org/abs/2505.11124} {arXiv:2505.11124 [physics.chem-ph]}
  \BibitemShut {NoStop}%
\bibitem [{\citenamefont {Mead}(1992{\natexlab{b}})}]{MeadRevModPhys.64.51}%
  \BibitemOpen
  \bibfield  {author} {\bibinfo {author} {\bibfnamefont {C.~A.}\ \bibnamefont
  {Mead}},\ }\bibfield  {title} {\enquote {\bibinfo {title} {The geometric
  phase in molecular systems},}\ }\href
  {https://doi.org/10.1103/RevModPhys.64.51} {\bibfield  {journal} {\bibinfo
  {journal} {Rev. Mod. Phys.}\ }\textbf {\bibinfo {volume} {64}},\ \bibinfo
  {pages} {51--85} (\bibinfo {year} {1992}{\natexlab{b}})}\BibitemShut
  {NoStop}%
\bibitem [{\citenamefont {Wittig}(2012)}]{wittig2012}%
  \BibitemOpen
  \bibfield  {author} {\bibinfo {author} {\bibfnamefont {C.}~\bibnamefont
  {Wittig}},\ }\bibfield  {title} {\enquote {\bibinfo {title} {Geometric phase
  and gauge connection in polyatomic molecules},}\ }\href
  {https://doi.org/10.1039/C2CP22974A} {\bibfield  {journal} {\bibinfo
  {journal} {Phys. Chem. Chem. Phys.}\ }\textbf {\bibinfo {volume} {14}},\
  \bibinfo {pages} {6409--6432} (\bibinfo {year} {2012})}\BibitemShut {NoStop}%
\bibitem [{\citenamefont {Yuan}\ \emph {et~al.}(2018)\citenamefont {Yuan},
  \citenamefont {Guan}, \citenamefont {Chen}, \citenamefont {Zhao},
  \citenamefont {Yu}, \citenamefont {Luo}, \citenamefont {Tan}, \citenamefont
  {Xie}, \citenamefont {Wang}, \citenamefont {Sun}, \citenamefont {Zhang},\
  and\ \citenamefont {Yang}}]{yuan2018a}%
  \BibitemOpen
  \bibfield  {author} {\bibinfo {author} {\bibfnamefont {D.}~\bibnamefont
  {Yuan}}, \bibinfo {author} {\bibfnamefont {Y.}~\bibnamefont {Guan}}, \bibinfo
  {author} {\bibfnamefont {W.}~\bibnamefont {Chen}}, \bibinfo {author}
  {\bibfnamefont {H.}~\bibnamefont {Zhao}}, \bibinfo {author} {\bibfnamefont
  {S.}~\bibnamefont {Yu}}, \bibinfo {author} {\bibfnamefont {C.}~\bibnamefont
  {Luo}}, \bibinfo {author} {\bibfnamefont {Y.}~\bibnamefont {Tan}}, \bibinfo
  {author} {\bibfnamefont {T.}~\bibnamefont {Xie}}, \bibinfo {author}
  {\bibfnamefont {X.}~\bibnamefont {Wang}}, \bibinfo {author} {\bibfnamefont
  {Z.}~\bibnamefont {Sun}}, \bibinfo {author} {\bibfnamefont {D.~H.}\
  \bibnamefont {Zhang}},\ and\ \bibinfo {author} {\bibfnamefont
  {X.}~\bibnamefont {Yang}},\ }\bibfield  {title} {\enquote {\bibinfo {title}
  {Observation of the geometric phase effect in the {{H}} + {{HD}}
  {$\rightarrow$} {{H2}} + {{D}} reaction},}\ }\href
  {https://doi.org/10.1126/science.aav1356} {\bibfield  {journal} {\bibinfo
  {journal} {Science}\ }\textbf {\bibinfo {volume} {362}},\ \bibinfo {pages}
  {1289--1293} (\bibinfo {year} {2018})}\BibitemShut {NoStop}%
\bibitem [{\citenamefont {Ryabinkin}, \citenamefont {{Joubert-Doriol}},\ and\
  \citenamefont {Izmaylov}(2017)}]{ryabinkin2017}%
  \BibitemOpen
  \bibfield  {author} {\bibinfo {author} {\bibfnamefont {I.~G.}\ \bibnamefont
  {Ryabinkin}}, \bibinfo {author} {\bibfnamefont {L.}~\bibnamefont
  {{Joubert-Doriol}}},\ and\ \bibinfo {author} {\bibfnamefont {A.~F.}\
  \bibnamefont {Izmaylov}},\ }\bibfield  {title} {\enquote {\bibinfo {title}
  {Geometric {{Phase Effects}} in {{Nonadiabatic Dynamics}} near {{Conical
  Intersections}}},}\ }\href {https://doi.org/10.1021/acs.accounts.7b00220}
  {\bibfield  {journal} {\bibinfo  {journal} {Acc. Chem. Res.}\ }\textbf
  {\bibinfo {volume} {50}},\ \bibinfo {pages} {1785--1793} (\bibinfo {year}
  {2017})}\BibitemShut {NoStop}%
\bibitem [{\citenamefont {Yan}\ and\ \citenamefont
  {Mukamel}(1988)}]{Yan885160}%
  \BibitemOpen
  \bibfield  {author} {\bibinfo {author} {\bibfnamefont {Y.}~\bibnamefont
  {Yan}}\ and\ \bibinfo {author} {\bibfnamefont {S.}~\bibnamefont {Mukamel}},\
  }\bibfield  {title} {\enquote {\bibinfo {title} {Electronic dephasing,
  vibrational relaxation, and solvent friction in molecular nonlinear optical
  lineshapes},}\ }\href {https://doi.org/10.1063/1.455632} {\bibfield
  {journal} {\bibinfo  {journal} {The Journal of Chemical Physics}\ }\textbf
  {\bibinfo {volume} {89}},\ \bibinfo {pages} {5160--5176} (\bibinfo {year}
  {1988})}\BibitemShut {NoStop}%
\bibitem [{\citenamefont {Nitzan}(2024)}]{nitzan_2024}%
  \BibitemOpen
  \bibfield  {author} {\bibinfo {author} {\bibfnamefont {A.}~\bibnamefont
  {Nitzan}},\ }\href {https://doi.org/10.1093/9780191947971.001.0001} {\emph
  {\bibinfo {title} {Chemical Dynamics in Condensed Phases: Relaxation,
  Transfer, and Reactions in Condensed Molecular Systems}}},\ \bibinfo
  {edition} {online edn}\ ed.\ (\bibinfo  {publisher} {Oxford University
  Press},\ \bibinfo {year} {2024})\ \bibinfo {note} {accessed:
  2025-08-25}\BibitemShut {NoStop}%
\bibitem [{\citenamefont {Kühl}\ and\ \citenamefont
  {Domcke}(2000)}]{KUHL2000227}%
  \BibitemOpen
  \bibfield  {author} {\bibinfo {author} {\bibfnamefont {A.}~\bibnamefont
  {Kühl}}\ and\ \bibinfo {author} {\bibfnamefont {W.}~\bibnamefont {Domcke}},\
  }\bibfield  {title} {\enquote {\bibinfo {title} {Effect of a dissipative
  environment on the dynamics at a conical intersection},}\ }\href
  {https://doi.org/https://doi.org/10.1016/S0301-0104(00)00199-3} {\bibfield
  {journal} {\bibinfo  {journal} {Chemical Physics}\ }\textbf {\bibinfo
  {volume} {259}},\ \bibinfo {pages} {227--236} (\bibinfo {year}
  {2000})}\BibitemShut {NoStop}%
\bibitem [{\citenamefont {Chen}\ \emph {et~al.}(2016)\citenamefont {Chen},
  \citenamefont {Gelin}, \citenamefont {Chernyak}, \citenamefont {Domcke},\
  and\ \citenamefont {Zhao}}]{C6FD00088F}%
  \BibitemOpen
  \bibfield  {author} {\bibinfo {author} {\bibfnamefont {L.}~\bibnamefont
  {Chen}}, \bibinfo {author} {\bibfnamefont {M.~F.}\ \bibnamefont {Gelin}},
  \bibinfo {author} {\bibfnamefont {V.~Y.}\ \bibnamefont {Chernyak}}, \bibinfo
  {author} {\bibfnamefont {W.}~\bibnamefont {Domcke}},\ and\ \bibinfo {author}
  {\bibfnamefont {Y.}~\bibnamefont {Zhao}},\ }\bibfield  {title} {\enquote
  {\bibinfo {title} {Dissipative dynamics at conical intersections: simulations
  with the hierarchy equations of motion method},}\ }\href
  {https://doi.org/10.1039/C6FD00088F} {\bibfield  {journal} {\bibinfo
  {journal} {Faraday Discuss.}\ }\textbf {\bibinfo {volume} {194}},\ \bibinfo
  {pages} {61--80} (\bibinfo {year} {2016})}\BibitemShut {NoStop}%
\bibitem [{\citenamefont {Tanimura}(2012)}]{Tan1222A550}%
  \BibitemOpen
  \bibfield  {author} {\bibinfo {author} {\bibfnamefont {Y.}~\bibnamefont
  {Tanimura}},\ }\bibfield  {title} {\enquote {\bibinfo {title} {Reduced
  hierarchy equations of motion approach with drude plus brownian spectral
  distribution: Probing electron transfer processes by means of two-dimensional
  correlation spectroscopy},}\ }\href {https://doi.org/10.1063/1.4766931}
  {\bibfield  {journal} {\bibinfo  {journal} {The Journal of Chemical Physics}\
  }\textbf {\bibinfo {volume} {137}},\ \bibinfo {pages} {22A550} (\bibinfo
  {year} {2012})},\ \Eprint
  {https://arxiv.org/abs/https://pubs.aip.org/aip/jcp/article-pdf/doi/10.1063/1.4766931/14007229/22a550\_1\_online.pdf}
  {https://pubs.aip.org/aip/jcp/article-pdf/doi/10.1063/1.4766931/14007229/22a550\_1\_online.pdf}
  \BibitemShut {NoStop}%
\bibitem [{\citenamefont {Duan}\ and\ \citenamefont
  {Thorwart}(2016)}]{Duan16382}%
  \BibitemOpen
  \bibfield  {author} {\bibinfo {author} {\bibfnamefont {H.-G.}\ \bibnamefont
  {Duan}}\ and\ \bibinfo {author} {\bibfnamefont {M.}~\bibnamefont
  {Thorwart}},\ }\bibfield  {title} {\enquote {\bibinfo {title} {Quantum
  mechanical wave packet dynamics at a conical intersection with strong
  vibrational dissipation},}\ }\href
  {https://doi.org/10.1021/acs.jpclett.5b02793} {\bibfield  {journal} {\bibinfo
   {journal} {The Journal of Physical Chemistry Letters}\ }\textbf {\bibinfo
  {volume} {7}},\ \bibinfo {pages} {382--386} (\bibinfo {year} {2016})},\
  \bibinfo {note} {pMID: 26751091},\ \Eprint
  {https://arxiv.org/abs/https://doi.org/10.1021/acs.jpclett.5b02793}
  {https://doi.org/10.1021/acs.jpclett.5b02793} \BibitemShut {NoStop}%
\bibitem [{\citenamefont {Ikeda}\ and\ \citenamefont
  {Tanimura}(2018)}]{IKEDA2018203}%
  \BibitemOpen
  \bibfield  {author} {\bibinfo {author} {\bibfnamefont {T.}~\bibnamefont
  {Ikeda}}\ and\ \bibinfo {author} {\bibfnamefont {Y.}~\bibnamefont
  {Tanimura}},\ }\bibfield  {title} {\enquote {\bibinfo {title} {Phase-space
  wavepacket dynamics of internal conversion via conical intersection:
  Multi-state quantum fokker-planck equation approach},}\ }\href
  {https://doi.org/https://doi.org/10.1016/j.chemphys.2018.07.013} {\bibfield
  {journal} {\bibinfo  {journal} {Chemical Physics}\ }\textbf {\bibinfo
  {volume} {515}},\ \bibinfo {pages} {203--213} (\bibinfo {year} {2018})},\
  \bibinfo {note} {ultrafast Photoinduced Processes in Polyatomic
  Molecules:Electronic Structure, Dynamics and Spectroscopy (Dedicated to
  Wolfgang Domcke on the occasion of his 70th birthday)}\BibitemShut {NoStop}%
\bibitem [{\citenamefont {Feynman}\ and\ \citenamefont {\mbox{Vernon,
  Jr.}}(1963)}]{Fey63118}%
  \BibitemOpen
  \bibfield  {author} {\bibinfo {author} {\bibfnamefont {R.~P.}\ \bibnamefont
  {Feynman}}\ and\ \bibinfo {author} {\bibfnamefont {F.~L.}\ \bibnamefont
  {\mbox{Vernon, Jr.}}},\ }\bibfield  {title} {\enquote {\bibinfo {title} {The
  theory of a general quantum system interacting with a linear dissipative
  system},}\ }\href@noop {} {\bibfield  {journal} {\bibinfo  {journal} {Ann.
  Phys.}\ }\textbf {\bibinfo {volume} {24}},\ \bibinfo {pages} {118--173}
  (\bibinfo {year} {1963})}\BibitemShut {NoStop}%
\bibitem [{\citenamefont {Gu}(2023)}]{gu2023}%
  \BibitemOpen
  \bibfield  {author} {\bibinfo {author} {\bibfnamefont {B.}~\bibnamefont
  {Gu}},\ }\bibfield  {title} {\enquote {\bibinfo {title} {A
  {{Discrete-Variable Local Diabatic Representation}} of {{Conical Intersection
  Dynamics}}},}\ }\href {https://doi.org/10.1021/acs.jctc.3c00560} {\bibfield
  {journal} {\bibinfo  {journal} {J. Chem. Theory Comput.}\ }\textbf {\bibinfo
  {volume} {19}},\ \bibinfo {pages} {6557--6563} (\bibinfo {year}
  {2023})}\BibitemShut {NoStop}%
\bibitem [{\citenamefont {Gu}(2024)}]{gu2024a}%
  \BibitemOpen
  \bibfield  {author} {\bibinfo {author} {\bibfnamefont {B.}~\bibnamefont
  {Gu}},\ }\bibfield  {title} {\enquote {\bibinfo {title} {Nonadiabatic
  {{Conical Intersection Dynamics}} in the {{Local Diabatic Representation}}
  with {{Strang Splitting}} and {{Fourier Basis}}},}\ }\href
  {https://doi.org/10.1021/acs.jctc.3c01317} {\bibfield  {journal} {\bibinfo
  {journal} {J. Chem. Theory Comput.}\ }\textbf {\bibinfo {volume} {20}},\
  \bibinfo {pages} {2711--2718} (\bibinfo {year} {2024})}\BibitemShut {NoStop}%
\bibitem [{\citenamefont {Tanimura}\ and\ \citenamefont
  {Kubo}(1989)}]{tanimura1989}%
  \BibitemOpen
  \bibfield  {author} {\bibinfo {author} {\bibfnamefont {Y.}~\bibnamefont
  {Tanimura}}\ and\ \bibinfo {author} {\bibfnamefont {R.}~\bibnamefont
  {Kubo}},\ }\bibfield  {title} {\enquote {\bibinfo {title} {Time {{Evolution}}
  of a {{Quantum System}} in {{Contact}} with a {{Nearly Gaussian-Markoffian
  Noise Bath}}},}\ }\href {https://doi.org/10.1143/JPSJ.58.101} {\bibfield
  {journal} {\bibinfo  {journal} {J. Phys. Soc. Jpn.}\ }\textbf {\bibinfo
  {volume} {58}},\ \bibinfo {pages} {101--114} (\bibinfo {year}
  {1989})}\BibitemShut {NoStop}%
\bibitem [{\citenamefont {Tanimura}(1990)}]{tanimura1990}%
  \BibitemOpen
  \bibfield  {author} {\bibinfo {author} {\bibfnamefont {Y.}~\bibnamefont
  {Tanimura}},\ }\bibfield  {title} {\enquote {\bibinfo {title}
  {Nonperturbative expansion method for a quantum system coupled to a
  harmonic-oscillator bath},}\ }\href
  {https://doi.org/10.1103/PhysRevA.41.6676} {\bibfield  {journal} {\bibinfo
  {journal} {Phys. Rev. A}\ }\textbf {\bibinfo {volume} {41}},\ \bibinfo
  {pages} {6676--6687} (\bibinfo {year} {1990})}\BibitemShut {NoStop}%
\bibitem [{\citenamefont {Xu}\ \emph {et~al.}(2005)\citenamefont {Xu},
  \citenamefont {Cui}, \citenamefont {Li}, \citenamefont {Mo},\ and\
  \citenamefont {Yan}}]{Xu05041103}%
  \BibitemOpen
  \bibfield  {author} {\bibinfo {author} {\bibfnamefont {R.~X.}\ \bibnamefont
  {Xu}}, \bibinfo {author} {\bibfnamefont {P.}~\bibnamefont {Cui}}, \bibinfo
  {author} {\bibfnamefont {X.~Q.}\ \bibnamefont {Li}}, \bibinfo {author}
  {\bibfnamefont {Y.}~\bibnamefont {Mo}},\ and\ \bibinfo {author}
  {\bibfnamefont {Y.~J.}\ \bibnamefont {Yan}},\ }\bibfield  {title} {\enquote
  {\bibinfo {title} {Exact quantum master equation via the calculus on path
  integrals},}\ }\href@noop {} {\bibfield  {journal} {\bibinfo  {journal} {J.
  Chem. Phys.}\ }\textbf {\bibinfo {volume} {122}},\ \bibinfo {pages} {041103}
  (\bibinfo {year} {2005})}\BibitemShut {NoStop}%
\bibitem [{\citenamefont {Xu}\ and\ \citenamefont {Yan}(2007)}]{Xu07031107}%
  \BibitemOpen
  \bibfield  {author} {\bibinfo {author} {\bibfnamefont {R.~X.}\ \bibnamefont
  {Xu}}\ and\ \bibinfo {author} {\bibfnamefont {Y.~J.}\ \bibnamefont {Yan}},\
  }\bibfield  {title} {\enquote {\bibinfo {title} {Dynamics of quantum
  dissipation systems interacting with bosonic canonical bath: Hierarchical
  equations of motion approach},}\ }\href@noop {} {\bibfield  {journal}
  {\bibinfo  {journal} {Phys. Rev. E}\ }\textbf {\bibinfo {volume} {75}},\
  \bibinfo {pages} {031107} (\bibinfo {year} {2007})}\BibitemShut {NoStop}%
\bibitem [{\citenamefont {Jin}, \citenamefont {Zheng},\ and\ \citenamefont
  {Yan}(2008)}]{jin2008}%
  \BibitemOpen
  \bibfield  {author} {\bibinfo {author} {\bibfnamefont {J.}~\bibnamefont
  {Jin}}, \bibinfo {author} {\bibfnamefont {X.}~\bibnamefont {Zheng}},\ and\
  \bibinfo {author} {\bibfnamefont {Y.}~\bibnamefont {Yan}},\ }\bibfield
  {title} {\enquote {\bibinfo {title} {Exact dynamics of dissipative electronic
  systems and quantum transport: {{Hierarchical}} equations of motion
  approach},}\ }\href {https://doi.org/10.1063/1.2938087} {\bibfield  {journal}
  {\bibinfo  {journal} {J. Chem. Phys.}\ }\textbf {\bibinfo {volume} {128}},\
  \bibinfo {pages} {234703} (\bibinfo {year} {2008})}\BibitemShut {NoStop}%
\bibitem [{\citenamefont {Yan}(2014)}]{yan2014}%
  \BibitemOpen
  \bibfield  {author} {\bibinfo {author} {\bibfnamefont {Y.}~\bibnamefont
  {Yan}},\ }\bibfield  {title} {\enquote {\bibinfo {title} {Theory of open
  quantum systems with bath of electrons and phonons and spins:
  {{Many-dissipaton}} density matrixes approach},}\ }\href
  {https://doi.org/10.1063/1.4863379} {\bibfield  {journal} {\bibinfo
  {journal} {J. Chem. Phys.}\ }\textbf {\bibinfo {volume} {140}},\ \bibinfo
  {pages} {054105} (\bibinfo {year} {2014})}\BibitemShut {NoStop}%
\bibitem [{\citenamefont {Yan}\ \emph {et~al.}(2004)\citenamefont {Yan},
  \citenamefont {Yang}, \citenamefont {Liu},\ and\ \citenamefont
  {Shao}}]{Yan04216}%
  \BibitemOpen
  \bibfield  {author} {\bibinfo {author} {\bibfnamefont {Y.~A.}\ \bibnamefont
  {Yan}}, \bibinfo {author} {\bibfnamefont {F.}~\bibnamefont {Yang}}, \bibinfo
  {author} {\bibfnamefont {Y.}~\bibnamefont {Liu}},\ and\ \bibinfo {author}
  {\bibfnamefont {J.~S.}\ \bibnamefont {Shao}},\ }\bibfield  {title} {\enquote
  {\bibinfo {title} {Hierarchical approach based on stochastic decoupling to
  dissipative systems},}\ }\href@noop {} {\bibfield  {journal} {\bibinfo
  {journal} {Chem. Phys. Lett.}\ }\textbf {\bibinfo {volume} {395}},\ \bibinfo
  {pages} {216--21} (\bibinfo {year} {2004})}\BibitemShut {NoStop}%
\bibitem [{\citenamefont {Xie}, \citenamefont {Yarkony},\ and\ \citenamefont
  {Guo}(2017)}]{xie2017b}%
  \BibitemOpen
  \bibfield  {author} {\bibinfo {author} {\bibfnamefont {C.}~\bibnamefont
  {Xie}}, \bibinfo {author} {\bibfnamefont {D.~R.}\ \bibnamefont {Yarkony}},\
  and\ \bibinfo {author} {\bibfnamefont {H.}~\bibnamefont {Guo}},\ }\bibfield
  {title} {\enquote {\bibinfo {title} {Nonadiabatic tunneling via conical
  intersections and the role of the geometric phase},}\ }\href
  {https://doi.org/10.1103/PhysRevA.95.022104} {\bibfield  {journal} {\bibinfo
  {journal} {Phys. Rev. A}\ }\textbf {\bibinfo {volume} {95}},\ \bibinfo
  {pages} {022104} (\bibinfo {year} {2017})}\BibitemShut {NoStop}%
\bibitem [{\citenamefont {Weimer}, \citenamefont {Kshetrimayum},\ and\
  \citenamefont {Or\'us}(2021)}]{Wei21015008}%
  \BibitemOpen
  \bibfield  {author} {\bibinfo {author} {\bibfnamefont {H.}~\bibnamefont
  {Weimer}}, \bibinfo {author} {\bibfnamefont {A.}~\bibnamefont
  {Kshetrimayum}},\ and\ \bibinfo {author} {\bibfnamefont {R.}~\bibnamefont
  {Or\'us}},\ }\bibfield  {title} {\enquote {\bibinfo {title} {Simulation
  methods for open quantum many-body systems},}\ }\href
  {https://doi.org/10.1103/RevModPhys.93.015008} {\bibfield  {journal}
  {\bibinfo  {journal} {Rev. Mod. Phys.}\ }\textbf {\bibinfo {volume} {93}},\
  \bibinfo {pages} {015008} (\bibinfo {year} {2021})}\BibitemShut {NoStop}%
\bibitem [{\citenamefont {Tanimura}(2020)}]{tanimura2020}%
  \BibitemOpen
  \bibfield  {author} {\bibinfo {author} {\bibfnamefont {Y.}~\bibnamefont
  {Tanimura}},\ }\bibfield  {title} {\enquote {\bibinfo {title} {Perspective:
  {{Numerically}} "exact" approach to open quantum dynamics: {{The}}
  hierarchical equations of motion ({{HEOM}})},}\ }\href
  {https://doi.org/10.1063/5.0011599} {\bibfield  {journal} {\bibinfo
  {journal} {J. Chem. Phys.}\ }\textbf {\bibinfo {volume} {153}},\ \bibinfo
  {pages} {020901} (\bibinfo {year} {2020})},\ \Eprint
  {https://arxiv.org/abs/2006.05501} {arXiv:2006.05501 [cond-mat,
  physics:physics, physics:quant-ph]} \BibitemShut {NoStop}%
\bibitem [{\citenamefont {Wang}\ and\ \citenamefont {Yan}(2022)}]{wang2022e}%
  \BibitemOpen
  \bibfield  {author} {\bibinfo {author} {\bibfnamefont {Y.}~\bibnamefont
  {Wang}}\ and\ \bibinfo {author} {\bibfnamefont {Y.}~\bibnamefont {Yan}},\
  }\bibfield  {title} {\enquote {\bibinfo {title} {Quantum mechanics of open
  systems: {{Dissipaton}} theories},}\ }\href
  {https://doi.org/10.1063/5.0123999} {\bibfield  {journal} {\bibinfo
  {journal} {J. Chem. Phys.}\ }\textbf {\bibinfo {volume} {157}},\ \bibinfo
  {pages} {170901} (\bibinfo {year} {2022})}\BibitemShut {NoStop}%
\bibitem [{\citenamefont {Hu}, \citenamefont {Xu},\ and\ \citenamefont
  {Yan}(2010)}]{Hu10101106}%
  \BibitemOpen
  \bibfield  {author} {\bibinfo {author} {\bibfnamefont {J.}~\bibnamefont
  {Hu}}, \bibinfo {author} {\bibfnamefont {R.~X.}\ \bibnamefont {Xu}},\ and\
  \bibinfo {author} {\bibfnamefont {Y.~J.}\ \bibnamefont {Yan}},\ }\bibfield
  {title} {\enquote {\bibinfo {title} {Pad\'{e} spectrum decomposition of fermi
  function and bose function},}\ }\href@noop {} {\bibfield  {journal} {\bibinfo
   {journal} {J. Chem. Phys.}\ }\textbf {\bibinfo {volume} {133}},\ \bibinfo
  {pages} {101106} (\bibinfo {year} {2010})}\BibitemShut {NoStop}%
\bibitem [{\citenamefont {Hu}\ \emph {et~al.}(2011)\citenamefont {Hu},
  \citenamefont {Luo}, \citenamefont {Jiang}, \citenamefont {Xu},\ and\
  \citenamefont {Yan}}]{Hu11244106}%
  \BibitemOpen
  \bibfield  {author} {\bibinfo {author} {\bibfnamefont {J.}~\bibnamefont
  {Hu}}, \bibinfo {author} {\bibfnamefont {M.}~\bibnamefont {Luo}}, \bibinfo
  {author} {\bibfnamefont {F.}~\bibnamefont {Jiang}}, \bibinfo {author}
  {\bibfnamefont {R.~X.}\ \bibnamefont {Xu}},\ and\ \bibinfo {author}
  {\bibfnamefont {Y.~J.}\ \bibnamefont {Yan}},\ }\bibfield  {title} {\enquote
  {\bibinfo {title} {Pad\'{e} spectrum decompositions of quantum distribution
  functions and optimal hierarchial equations of motion construction for
  quantum open systems},}\ }\href@noop {} {\bibfield  {journal} {\bibinfo
  {journal} {J. Chem. Phys.}\ }\textbf {\bibinfo {volume} {134}},\ \bibinfo
  {pages} {244106} (\bibinfo {year} {2011})}\BibitemShut {NoStop}%
\bibitem [{\citenamefont {Ding}\ \emph {et~al.}(2011)\citenamefont {Ding},
  \citenamefont {Xu}, \citenamefont {Hu}, \citenamefont {Xu},\ and\
  \citenamefont {Yan}}]{Din11164107}%
  \BibitemOpen
  \bibfield  {author} {\bibinfo {author} {\bibfnamefont {J.~J.}\ \bibnamefont
  {Ding}}, \bibinfo {author} {\bibfnamefont {J.}~\bibnamefont {Xu}}, \bibinfo
  {author} {\bibfnamefont {J.}~\bibnamefont {Hu}}, \bibinfo {author}
  {\bibfnamefont {R.~X.}\ \bibnamefont {Xu}},\ and\ \bibinfo {author}
  {\bibfnamefont {Y.~J.}\ \bibnamefont {Yan}},\ }\bibfield  {title} {\enquote
  {\bibinfo {title} {Optimized hierarchical equations of motion for drude
  dissipation with applications to linear and nonlinear optical responses},}\
  }\href@noop {} {\bibfield  {journal} {\bibinfo  {journal} {J. Chem. Phys.}\
  }\textbf {\bibinfo {volume} {135}},\ \bibinfo {pages} {164107} (\bibinfo
  {year} {2011})}\BibitemShut {NoStop}%
\bibitem [{\citenamefont {Ding}, \citenamefont {Xu},\ and\ \citenamefont
  {Yan}(2012)}]{Din12224103}%
  \BibitemOpen
  \bibfield  {author} {\bibinfo {author} {\bibfnamefont {J.~J.}\ \bibnamefont
  {Ding}}, \bibinfo {author} {\bibfnamefont {R.~X.}\ \bibnamefont {Xu}},\ and\
  \bibinfo {author} {\bibfnamefont {Y.~J.}\ \bibnamefont {Yan}},\ }\bibfield
  {title} {\enquote {\bibinfo {title} {Optimizing hierarchical equations of
  motion for quantum dissipation and quantifying quantum bath effects on
  quantum transfer mechanisms},}\ }\href@noop {} {\bibfield  {journal}
  {\bibinfo  {journal} {J. Chem. Phys.}\ }\textbf {\bibinfo {volume} {136}},\
  \bibinfo {pages} {224103} (\bibinfo {year} {2012})}\BibitemShut {NoStop}%
\bibitem [{\citenamefont {Zheng}\ \emph {et~al.}(2012)\citenamefont {Zheng},
  \citenamefont {Xu}, \citenamefont {Xu}, \citenamefont {Jin}, \citenamefont
  {Hu},\ and\ \citenamefont {Yan}}]{Zhe121129}%
  \BibitemOpen
  \bibfield  {author} {\bibinfo {author} {\bibfnamefont {X.}~\bibnamefont
  {Zheng}}, \bibinfo {author} {\bibfnamefont {R.~X.}\ \bibnamefont {Xu}},
  \bibinfo {author} {\bibfnamefont {J.}~\bibnamefont {Xu}}, \bibinfo {author}
  {\bibfnamefont {J.~S.}\ \bibnamefont {Jin}}, \bibinfo {author} {\bibfnamefont
  {J.}~\bibnamefont {Hu}},\ and\ \bibinfo {author} {\bibfnamefont {Y.~J.}\
  \bibnamefont {Yan}},\ }\bibfield  {title} {\enquote {\bibinfo {title}
  {Hierarchical equations of motion for quantum dissipation and quantum
  transport},}\ }\href@noop {} {\bibfield  {journal} {\bibinfo  {journal}
  {Prog. Chem.}\ }\textbf {\bibinfo {volume} {24}},\ \bibinfo {pages}
  {1129--1152} (\bibinfo {year} {2012})}\BibitemShut {NoStop}%
\bibitem [{\citenamefont {Chen}\ \emph {et~al.}(2022)\citenamefont {Chen},
  \citenamefont {Wang}, \citenamefont {Zheng}, \citenamefont {Xu},\ and\
  \citenamefont {Yan}}]{Che22221102}%
  \BibitemOpen
  \bibfield  {author} {\bibinfo {author} {\bibfnamefont {Z.~H.}\ \bibnamefont
  {Chen}}, \bibinfo {author} {\bibfnamefont {Y.}~\bibnamefont {Wang}}, \bibinfo
  {author} {\bibfnamefont {X.}~\bibnamefont {Zheng}}, \bibinfo {author}
  {\bibfnamefont {R.~X.}\ \bibnamefont {Xu}},\ and\ \bibinfo {author}
  {\bibfnamefont {Y.~J.}\ \bibnamefont {Yan}},\ }\bibfield  {title} {\enquote
  {\bibinfo {title} {Universal time-domain prony fitting decomposition for
  optimized hierarchical quantum master equations},}\ }\href@noop {} {\bibfield
   {journal} {\bibinfo  {journal} {J. Chem. Phys.}\ }\textbf {\bibinfo {volume}
  {156}},\ \bibinfo {pages} {221102} (\bibinfo {year} {2022})}\BibitemShut
  {NoStop}%
\bibitem [{\citenamefont {Takahashi}\ \emph {et~al.}(2024)\citenamefont
  {Takahashi}, \citenamefont {Rudge}, \citenamefont {Kaspar}, \citenamefont
  {Thoss},\ and\ \citenamefont {Borrelli}}]{Tak24204105}%
  \BibitemOpen
  \bibfield  {author} {\bibinfo {author} {\bibfnamefont {H.}~\bibnamefont
  {Takahashi}}, \bibinfo {author} {\bibfnamefont {S.}~\bibnamefont {Rudge}},
  \bibinfo {author} {\bibfnamefont {C.}~\bibnamefont {Kaspar}}, \bibinfo
  {author} {\bibfnamefont {M.}~\bibnamefont {Thoss}},\ and\ \bibinfo {author}
  {\bibfnamefont {R.}~\bibnamefont {Borrelli}},\ }\bibfield  {title} {\enquote
  {\bibinfo {title} {{High accuracy exponential decomposition of bath
  correlation functions for arbitrary and structured spectral densities:
  Emerging methodologies and new approaches}},}\ }\href@noop {} {\bibfield
  {journal} {\bibinfo  {journal} {J. Chem. Phys.}\ }\textbf {\bibinfo {volume}
  {160}},\ \bibinfo {pages} {204105} (\bibinfo {year} {2024})}\BibitemShut
  {NoStop}%
\bibitem [{\citenamefont {Yan}\ \emph {et~al.}(2016)\citenamefont {Yan},
  \citenamefont {Jin}, \citenamefont {Xu},\ and\ \citenamefont
  {Zheng}}]{Yan16110306}%
  \BibitemOpen
  \bibfield  {author} {\bibinfo {author} {\bibfnamefont {Y.~J.}\ \bibnamefont
  {Yan}}, \bibinfo {author} {\bibfnamefont {J.~S.}\ \bibnamefont {Jin}},
  \bibinfo {author} {\bibfnamefont {R.~X.}\ \bibnamefont {Xu}},\ and\ \bibinfo
  {author} {\bibfnamefont {X.}~\bibnamefont {Zheng}},\ }\bibfield  {title}
  {\enquote {\bibinfo {title} {Dissipaton equation of motion approach to open
  quantum systems},}\ }\href@noop {} {\bibfield  {journal} {\bibinfo  {journal}
  {Front. Phys.}\ }\textbf {\bibinfo {volume} {11}},\ \bibinfo {pages} {110306}
  (\bibinfo {year} {2016})}\BibitemShut {NoStop}%
\bibitem [{\citenamefont {Tanimura}(2006)}]{tanimura2006}%
  \BibitemOpen
  \bibfield  {author} {\bibinfo {author} {\bibfnamefont {Y.}~\bibnamefont
  {Tanimura}},\ }\bibfield  {title} {\enquote {\bibinfo {title} {Stochastic
  {{Liouville}}, {{Langevin}}, {{Fokker}}--{{Planck}}, and {{Master Equation
  Approaches}} to {{Quantum Dissipative Systems}}},}\ }\href
  {https://doi.org/10.1143/JPSJ.75.082001} {\bibfield  {journal} {\bibinfo
  {journal} {J. Phys. Soc. Jpn.}\ }\textbf {\bibinfo {volume} {75}},\ \bibinfo
  {pages} {082001} (\bibinfo {year} {2006})}\BibitemShut {NoStop}%
\bibitem [{\citenamefont {Zhu}\ \emph {et~al.}(2012)\citenamefont {Zhu},
  \citenamefont {Liu}, \citenamefont {Xie},\ and\ \citenamefont
  {Shi}}]{Zhu12194106}%
  \BibitemOpen
  \bibfield  {author} {\bibinfo {author} {\bibfnamefont {L.~L.}\ \bibnamefont
  {Zhu}}, \bibinfo {author} {\bibfnamefont {H.}~\bibnamefont {Liu}}, \bibinfo
  {author} {\bibfnamefont {W.~W.}\ \bibnamefont {Xie}},\ and\ \bibinfo {author}
  {\bibfnamefont {Q.}~\bibnamefont {Shi}},\ }\bibfield  {title} {\enquote
  {\bibinfo {title} {Explicit system-bath correlation calculated using the
  hierarchical equations of motion method},}\ }\href@noop {} {\bibfield
  {journal} {\bibinfo  {journal} {J. Chem. Phys.}\ }\textbf {\bibinfo {volume}
  {137}},\ \bibinfo {pages} {194106} (\bibinfo {year} {2012})}\BibitemShut
  {NoStop}%
\bibitem [{\citenamefont {Li}\ \emph {et~al.}(2023)\citenamefont {Li},
  \citenamefont {Su}, \citenamefont {Chen}, \citenamefont {Wang}, \citenamefont
  {Xu}, \citenamefont {Zheng},\ and\ \citenamefont {Yan}}]{Li_2023}%
  \BibitemOpen
  \bibfield  {author} {\bibinfo {author} {\bibfnamefont {X.}~\bibnamefont
  {Li}}, \bibinfo {author} {\bibfnamefont {Y.}~\bibnamefont {Su}}, \bibinfo
  {author} {\bibfnamefont {Z.-H.}\ \bibnamefont {Chen}}, \bibinfo {author}
  {\bibfnamefont {Y.}~\bibnamefont {Wang}}, \bibinfo {author} {\bibfnamefont
  {R.-X.}\ \bibnamefont {Xu}}, \bibinfo {author} {\bibfnamefont
  {X.}~\bibnamefont {Zheng}},\ and\ \bibinfo {author} {\bibfnamefont
  {Y.}~\bibnamefont {Yan}},\ }\bibfield  {title} {\enquote {\bibinfo {title}
  {Dissipatons as generalized brownian particles for open quantum systems:
  Dissipaton-embedded quantum master equation},}\ }\href
  {https://doi.org/10.1063/5.0151239} {\bibfield  {journal} {\bibinfo
  {journal} {The Journal of Chemical Physics}\ }\textbf {\bibinfo {volume}
  {158}} (\bibinfo {year} {2023}),\ 10.1063/5.0151239}\BibitemShut {NoStop}%
\bibitem [{\citenamefont {Zhang}\ and\ \citenamefont
  {Yan}(2015)}]{Zha15214112}%
  \BibitemOpen
  \bibfield  {author} {\bibinfo {author} {\bibfnamefont {H.~D.}\ \bibnamefont
  {Zhang}}\ and\ \bibinfo {author} {\bibfnamefont {Y.~J.}\ \bibnamefont
  {Yan}},\ }\bibfield  {title} {\enquote {\bibinfo {title} {Onsets of hierarchy
  truncation and self-consistent born approximation with quantum mechanics
  prescriptions invariance},}\ }\href@noop {} {\bibfield  {journal} {\bibinfo
  {journal} {J. Chem. Phys.}\ }\textbf {\bibinfo {volume} {143}},\ \bibinfo
  {pages} {214112} (\bibinfo {year} {2015})}\BibitemShut {NoStop}%
\bibitem [{\citenamefont {Zhang}\ \emph {et~al.}(2023)\citenamefont {Zhang},
  \citenamefont {Zuo}, \citenamefont {Ye}, \citenamefont {Chen}, \citenamefont
  {Wang}, \citenamefont {Xu}, \citenamefont {Zheng},\ and\ \citenamefont
  {Yan}}]{Zha23014106}%
  \BibitemOpen
  \bibfield  {author} {\bibinfo {author} {\bibfnamefont {D.}~\bibnamefont
  {Zhang}}, \bibinfo {author} {\bibfnamefont {L.}~\bibnamefont {Zuo}}, \bibinfo
  {author} {\bibfnamefont {L.}~\bibnamefont {Ye}}, \bibinfo {author}
  {\bibfnamefont {Z.-H.}\ \bibnamefont {Chen}}, \bibinfo {author}
  {\bibfnamefont {Y.}~\bibnamefont {Wang}}, \bibinfo {author} {\bibfnamefont
  {R.-X.}\ \bibnamefont {Xu}}, \bibinfo {author} {\bibfnamefont
  {X.}~\bibnamefont {Zheng}},\ and\ \bibinfo {author} {\bibfnamefont
  {Y.}~\bibnamefont {Yan}},\ }\bibfield  {title} {\enquote {\bibinfo {title}
  {{Hierarchical equations of motion approach for accurate characterization of
  spin excitations in quantum impurity systems}},}\ }\href@noop {} {\bibfield
  {journal} {\bibinfo  {journal} {J. Chem. Phys.}\ }\textbf {\bibinfo {volume}
  {158}},\ \bibinfo {pages} {014106} (\bibinfo {year} {2023})}\BibitemShut
  {NoStop}%
\end{thebibliography}%

\clearpage

\end{document}